\documentclass[fleqn]{aa}

\usepackage{amsmath}
\usepackage{natbib}
\usepackage{psfrag, color, epsfig, rotating, pstricks, pst-node}
\usepackage{dsfont}
\usepackage{txfonts}

\bibpunct{(}{)}{;}{a}{}{,}

\newlength{\hsizethird}
\newcommand{\dd}{{\mathrm d}}
\newcommand{\J}{{\mathrm J}}

\renewcommand{\vec}[1]{\boldsymbol{#1}}

\newcommand{\Omegam}{\Omega_{\rm m}}

\newcommand{\Omegade}{\Omega_{\rm de}}
\newcommand{\Omegab}{\Omega_{\rm b}}
\newcommand{\ns}{n_{\rm s}}

\begin{document}

\DeclareGraphicsExtensions{.eps, .ps}

\title{Dark-energy constraints and correlations with systematics from CFHTLS weak
  lensing, SNLS supernovae Ia and WMAP5
\thanks{Based on observations obtained with {\sc MegaPrime/MegaCam}, a
  joint project of CFHT and CEA/DAPNIA, at the Canada-France-Hawaii
  Telescope (CFHT) which is operated by the National Research Council
  (NRC) of Canada, the Institut National des Sciences de l'Univers of
  the Centre National de la Recherche Scientifique (CNRS) of France,
  and the University of Hawaii.  This work is based in part on data
  products produced at {\sc Terapix} and the Canadian Astronomy Data
  Centre as part of the Canada-France-Hawaii Telescope Legacy Survey,
  a collaborative project of NRC and CNRS.}
}

\titlerunning{Dark-energy constraints from weak lensing, SNIa and CMB}

\author{M.~Kilbinger\inst{1}
  \and K.~Benabed\inst{1}
  \and J.~Guy\inst{2}
  \and P.~Astier\inst{2}
  \and I.~Tereno\inst{1,3}
  \and L.~Fu\inst{4,1,5}
  \and D.~Wraith\inst{6,1}
  \and J.~Coupon\inst{1}
  \and Y.~Mellier\inst{1}
  \and C.~Balland\inst{2}
  \and F.~R.~Bouchet\inst{1}
  \and T.~Hamana\inst{7}
  \and D.~Hardin\inst{2}
  \and H.~J.~McCracken\inst{1}
  \and R.~Pain\inst{2}
  \and N.~Regnault\inst{2}
  \and M.~Schultheis\inst{8}
  \and H.~Yahagi\inst{9}
}

\institute{
  Institut d'Astrophysique de Paris, CNRS UMR 7095 \& UPMC, 98 bis,
  boulevard Arago, 75014 Paris, France
  \and
  LPNHE, CNRS-IN2P3 and Universit\'es Paris VI \& VII,
  4 place Jussieu, 75252 Paris Cedex 05, France
  \and
  Argelander-Institut f\"ur Astronomie, Universit\"at Bonn, Auf dem
  H\"ugel 71, 53121 Bonn, Germany
  \and
  INAF, Osservatorio Astronomico di Capodimonte, via Moiariello
  16, 80131 Napoli, Italy
  \and
  Shanghai Key Lab for Astrophysics, Shanghai Normal University,
  Shanghai 200234, P. R. China
  \and
  CEREMADE, Universit´e Paris Dauphine, 75775 Paris cedex 16, France
  \and
  National Astronomical Observatory of Japan, Mitaka, Tokyo
  181-8588, Japan
  \and
  Observatoire de Besan\c{c}on, 41bis, avenue de l'Observatoire, BP
  1615, 25010 Besan\c{c}on Cedex, France
  \and
  Research Institute for Information Technology, University of Kyushu
  6-10-1 Hakozaki, Higashi-ward, Fukuoka 812-8581, Japan
}

\offprints{Martin Kilbinger, \email{kilbinger@iap.fr}}

\date{Received / Accepted}

\abstract%
%
%
{}
%
%
{ We combine measurements of weak gravitational lensing from the
  CFHTLS-Wide survey, supernovae Ia from CFHT SNLS and CMB
  anisotropies from WMAP5 to obtain joint constraints on cosmological
  parameters, in particular, the dark-energy equation-of-state
  parameter $w$. We assess the influence of systematics in the data on
  the results and look for possible correlations with cosmological
  parameters.  } %
%
%
{ We implemented an MCMC algorithm to sample the parameter space of a
  flat CDM model with a dark-energy component of constant
  $w$. Systematics in the data are parametrised and included in the
  analysis. We determine the influence of photometric calibration of
  SNIa data on cosmological results by calculating the
  response of the distance modulus to photometric zero-point
  variations. The weak lensing data set is tested for anomalous
  field-to-field variations and a systematic shape measurement bias
  for high-redshift galaxies.}
%
%
%
{ Ignoring photometric uncertainties for SNLS biases cosmological
  parameters by at most 20\% of the statistical errors, using
  supernovae alone; the parameter uncertainties are underestimated by
  10\%. The weak-lensing field-to-field variance between 1
  deg$^2$-MegaCam pointings is 5\%-15\% higher than predicted
  from $N$-body simulations. We find no
  bias in the lensing signal at high redshift, within
  the framework of a simple model, and marginalising over cosmological
  parameters.
  Assuming a systematic underestimation of the lensing signal,
  the normalisation $\sigma_8$ increases by up to 8\%.  Combining all
  three probes we obtain $-0.10 < 1+w < 0.06$ at 68\% confidence
  ($-0.18<1+w<0.12$ at 95\%), including systematic errors. Our results
  are therefore consistent with the cosmological constant
  $\Lambda$. Systematics in the data increase the error bars by up to
  35\%; the best-fit values change by less than 0.15$\sigma$. }
%
%
{}


\maketitle

\section{Introduction}
\label{sec:intro}

The Canada-France-Hawaii-Telescope Legacy Survey
(CFHTLS\footnote{{http://www.cfht.hawaii.edu/Science/CFHTLS}})
has yielded interesting constraints on cosmological parameters using
different probes, for example weak gravitational lensing and
supernovae of Type Ia. In this paper we combine two recent
measurements from the CFHTLS: the cosmic shear data \citep[][hereafter
  F08]{FSHK08} and supernova (SN) Ia data from the Supernova Legacy
Survey \citep[SNLS,][hereafter A06]{2006A&A...447...31A}. We
complement these with CMB measurements from the Wilkinson Microwave
Anisotropy Probe \citep[WMAP5,][hereafter H09]{WMAP5-Hinshaw08}.

There are hundreds of type-Ia supernovae for which high-quality
observations are available. Many surveys spanning wide redshift ranges
and using different telescopes and search strategies can be
accessed. A recent compilation of various surveys and the derived
cosmological constraints can be found in
\citet{kowalski-2008}. Despite this impressive number of available
objects we have chosen to restrict ourselves to the SNLS sample. The
resulting constraints will suffer from a greater statistical
uncertainty; however, the systematic errors are better understood by
using a single, homogeneous survey. By providing imaging with a single
telescope and camera for example, SNLS allows a common photometric
calibration strategy for the whole survey.

The lensing data presented in F08 might suffer from systematic errors
in the shear calibration and shape measurement. In this work, we
performed two tests for the presence of systematics. First, we compared
the variance between individual MegaCam pointings to simulations.
Second, we quantified a potential systematic underestimation of the
lensing signal at high redshifts.
The influence on cosmological parameters was estimated. Due to the
unknown origin of those systematics, the analysis must remain less
rigorous than for the SNIa case. It is by no means comprehensive since
we only quantify the effect of those systematics and do not
investigate their origin, which is beyond the scope of this paper.

The analysis of the five-year WMAP data has yielded impressive results
on many cosmological parameters from a great number of models
\citep{WMAP5-Dunkley08, WMAP5-Hinshaw08, WMAP5-Komatsu08}. Thanks to
the rich features in the angular power spectrum, many cosmological
parameters can be determined with high precision. However,
degeneracies between parameters remain, in particular, for dark-energy
models.  In order to lift these degeneracies, measurements of CMB
anisotropies have to be complemented with other probes.

The WMAP5 team has combined their data with other probes which are
sensitive only to the geometry of the Universe, i.e.~SNIa and baryonic
acoustic oscillations (BAO). Other teams have included probes of
structure growth, like counts of X-ray clusters
\citep{2008MNRAS.387.1179M}, SDSS and Ly$\alpha$-forest
\citep{2008PhRvD..78h3524X}. In this work, we combine WMAP5 with weak
gravitational lensing which is sensitive to both the geometry and the
growth of structure. Probing both domains will allow future surveys
to distinguish between dark energy and modified gravity as a possible
cause for the present acceleration of the Universe
\cite[e.g.][]{2008PhRvD..78f3503J}. Such a test will be feasible with
upcoming and proposed surveys such as
KIDS\footnote{{http://www.astro-wise.org/projects/KIDS}},
DES\footnote{{https://www.darkenergysurvey.org}},
LSST\footnote{{www.lsst.org}},
JDEM\footnote{{http://universe.nasa.gov/program/probes/jdem.html}}
and
Euclid\footnote{
http://sci.esa.int/science-e/www/area/index.cfm?fareaid=102}.

The first-year WMAP data has already been combined with weak lensing,
using the RCS survey \citep{2002ApJ...577..595H}, to improve
constraints on $\Omegam$ and $\sigma_8$ \citep{Contaldi03}. In F08,
cosmic shear has been supplemented with the third-year WMAP data. In
this work we extend the latter, simple analysis by dropping the (not
well motivated) priors on some parameters for lensing ($\Omegab, \ns,
h$).

The `concordance' flat $\Lambda$CDM model of cosmology provides an
excellent fit to WMAP5 and most other probes of the geometry and
large-scale structure of the Universe. It only contains six free
parameters, $\Omegab$, $\Omegam$, $\tau$, $\ns$, $h$ and $\sigma_8$
(or functions thereof). This model assumes a cosmological constant
$\Lambda$ as the cause of the observed accelerated expansion of the
Universe today. The cosmological constant has yet evaded all plausible
physical explanations of its nature and origin and, further, brings in
problems of fine-tuning and coincidence.  A signature of some `dark
energy' beyond a simple cosmological constant might be an
equation-of-state (eos) parameter $w = p/\rho c^2$ which deviates from
the vacuum energy value of $-1$. We therefore extend the concordance
six-parameter model by including the dark-energy eos parameter
$w$. Although this $w$CDM  model with a constant $w$ is not better
motivated physically than the cosmological constant, a significant
observed deviation from $w=-1$ will definitely be an indication for
new physics. Moreover, the data at present are not good enough to
constrain more dark energy parameters in a general way. This is only
feasible for very specific models, e.g. quintessence
\citep[see][]{2007A&A...463..405S} or early dark energy
\citep{2006A&A...454...27B,2008arXiv0808.2840F}, and will be subject
of a future work using the CFHTLS data (Tereno et al.~in prep.).

One goal of this paper is to focus on systematic errors, nuisance
parameters and their interplay with cosmological parameters. For the
SNIa data, apart from the usual light-curve parameters, we take into
account photometric calibration errors. A joint Bayesian analysis
including systematic and cosmological parameters is done and
correlations are revealed. This will be mandatory for future surveys
with statistical errors which will be smaller by several orders of
magnitude as compared to today. To decrease systematics further and
further is very challenging, and technical limitations might set a
barrier to this endeavour.  Therefore, it is important to quantify the
effect of systematics and nuisance factors on cosmological
constraints.

This paper is organised as follows. Section \ref{sec:data} contains a
brief description of the data together with the likelihood used later
in the analysis. It also discusses sources of systematics in the data
and their correlations with cosmological parameters. In
Sect.~\ref{sec:constraints} we define the cosmological model tested in
this work, introduce our implementation of the Monte Carlo Markov
Chain (MCMC) technique and present the cosmological results of
the analysis. We conclude with a discussion in
Sect.~\ref{sec:discussion} and an outlook in Sect.~\ref{sec:outlook}.

\section{Data and method}
\label{sec:data}

\subsection{CFHTLS-Wide cosmic shear}
\label{sec:lensing-data}

We use the cosmic shear results from the CFHTLS-Wide $3^{\rm rd}$ year
data release \citep[T0003,][]{FSHK08}. On 57 square degrees (35
sq.~deg.~effective area), about $2\times 10^6$ galaxies with
$i_{AB}$-magnitudes between 21.5 and 24.5 were imaged. The data,
reduction analyses and shear pipeline are described in detail in F08.
We use the aperture-mass dispersion \citep{1998MNRAS.296..873S}
measured between 2 and 230 arc minutes. Due to the compensated nature
of its filter, this second-order measure is least sensitive to
large-scale systmatics in the data. The source redshift distribution
is obtained by using the CFHTLS-Deep $p(z)$
\citep{2006A&A...457..841I} and by rescaling it according to the Wide
$i_{AB}$ magnitue distribution and weak-lensing galaxy weights.

To ascertain that the quality and reliability of the shear
measurements are sufficient for this work, we perform further tests of
the data beyond what has been done in F08. This is addressed in
Sects.~\ref{sec:systematics-redshift} and
\ref{sec:systematics-variance} where we assess the importance of
potential systematics for the current data, and estimate their
influence on inferred cosmological parameters.

As in F08 the log-likelihood is $\ln L = - \chi^2/2 - \ln |C|/2 +
{\rm const}$, where the $\chi^2$ is modelled as
\begin{align}
  \chi^2_{\rm wl}(\vec p) = \sum_{ij} &
  \left( \left\langle M_{\rm ap}^2(\theta_i)
  \right\rangle_{\rm obs} -  \left\langle M_{\rm ap}^2(\theta_i, \vec p) \right\rangle
  \right) [C^{-1}]_{ij} \nonumber \\
  & \times 
  \left( \left\langle M_{\rm ap}^2(\theta_j)
  \right\rangle_{\rm obs} -  \left\langle M_{\rm ap}^2(\theta_j, \vec p) \right\rangle
  \right). 
    \label{chi2_lens}
\end{align}
The predicted aperture-mass dispersion given by a model parameter
vector $\vec p$ is fitted to $\langle M_{\rm ap}^2 \rangle_{\rm obs}$
measured at angular scales $\theta_i$.  The covariance $C$ of $\langle
M_{\rm ap}^2 \rangle_{\rm obs}$ is the one used in F08 and contains
shape noise, (non-Gaussian) cosmic variance and the residual
B-mode. Those parts of the covariance which depend on the shear
correlation (mixed and cosmic variance terms) are calculated using a
theoretical model of the large-scale structure and therefore depend on
cosmological parameters. We ignore this dependence and keep the
covariance constant, corresponding to the fiducial cosmology as in
F08. This biases the posterior confidence regions, but the effect is
weak over the region of parameter space permitted by CMB and lensing,
see \citet{ESH09} for a detailed discussion.  Moreover, we can drop
the term $\ln | C |/2$ in the log-likelihood because its
parameter-dependence manifests itself only for very small survey areas
\citep{KM05}.

As usual, the following relation between the aperture-mass dispersion
and the weak lensing power spectrum holds \citep{1998MNRAS.296..873S},
\begin{equation}
  \langle M_{\rm ap}^2 \rangle(\theta) = \int \frac{\dd \ell\,
    \ell}{2\pi} P_\kappa(\ell) \left[ \frac{24 \, \J_4(\theta
    \ell)}{(\theta \ell)^2}\right]^2 .
\end{equation}
The lensing power spectrum is a projection of the 3d matter-density
power spectrum $P_\delta$, weighted by the redshift distribution
$p(\chi)$ \citep{1992ApJ...388..272K},
\begin{align}
  P_\kappa(\ell) = & \frac 9 4 \Omegam^2 \left(\frac{H_0} c\right)^4
  \int_0^{\chi_{\rm lim}} \frac{\dd \chi}{a^2(\chi)}
  P_\delta(\ell/\chi; \chi) \nonumber \\
  & \times
  \left[
    \int_\chi^{\chi_{\rm lim}} \dd\chi^\prime p(\chi^\prime) \frac{\chi^\prime -
      \chi}{\chi^\prime} \right]^2.
\end{align}
For the non-linear evolution of the power spectrum, the fitting
formula of \citet{2003MNRAS.341.1311S} is used. Although tested for
$\Lambda$CDM models, it provides reasonable good fits to $w$CDM
cosmologies as well \citep{2007ApJ...665..887M}. The accuracy of any
non-linear fitting function is limited; when using the ansatz of
\cite{PD96} instead of \citet{2003MNRAS.341.1311S}, the resulting
best-fit $\sigma_8$ differs by 2\% (F08).

We parametrise the redshift distribution using the function
\begin{equation}
  p(z) \propto \frac{z^a + z^{ab}}{z^b + c} ; \;\;\;\; \int_0^{z_{\rm
      max}} p(z) \, \dd z = 1,
  \label{pofz}
\end{equation}
which we fit to the obtained redshift histogram. The
corresponding $\chi^2$ is
\begin{equation}
  \chi_{z}^2 =  \sum_i \frac{\left[ n_i -
      p(z_i)\right]^2}{\sigma_i^2}.
  \label{chi2nofz}
\end{equation}
Here, $n_i$ is the normalised number of galaxies in the $i^{\rm th}$
redshift bin and $p(z_i)$ the fitting function, evaluated at the
redshift bin centre.  The uncertainty $\sigma_i$ of $n_i$ contains
Poisson noise, photo-$z$ uncertainty and cosmic variance, as described
in F08 and \cite{JonBen07}. The sum in eq.~(\ref{chi2nofz})
extends over the redshift range $0 \le z \le 2.5$. In this range,
cosmic variance is the dominant uncertainty. As in F08 we neglect the
cross-correlation between different bins.

\subsubsection{Systematics in the lensing data}

In the following two subsections we address the question of potential
residual systematics in the cosmic shear data. We estimate the
influence of some systematics on cosmological constraints. The shear
catalogue used here has been extensively tested in F08. Internal
consistency checks have been performed involving the comparison
between two data reduction pipelines, residual B-modes and
cross-correlations between galaxies and stars. In addition, the shape
measurement pipeline has been calibrated with the STEP1 and STEP2
simulations \citep[][]{STEP1, STEP2}, and a bias of less than 3\% has
been determined.

Despite that, there are indications of remaining systematics which
do not manifest themselves in the merged shear catalogue. For example,
there are variations of the shear signal between individual MegaCam
pointings which vanish on average. Moreover, with the help of
photometric redshifts covering parts of the survey, a problem with the
redshift-scaling of the shear correlation becomes apparent.

\subsubsection{Variance between CFHTLS pointings}
\label{sec:systematics-variance}

In F08 it was shown the systematics are small globally. The
B-mode is consistent with zero on most scales (note however a
significant detection around 100 arc minutes). The level of
systematics is low on average, e.g.~the residual cross-correlation
between uncorrected star ellipticities and galaxy shapes after PSF
correction. There might however be systematics present on individual
fields. Even if they vanish on average they are an additional noise
source and increase the measurement error bars.

Here, we address the question whether individual pointings show
anomalous variations which might be due to incorrect shape
measurements or PSF correction. 
The field-to-field variance of the lensing signal is compared with
numerical simulations.

We measure the shear aperture-mass dispersion on all 57
fields, each one corresponding to a 1 deg$^2$-MegaCam image, and calculate
the field-to-field variance. We use $N$-body simulations
\citep{2001ApJ...558..463Y,2005PASJ...57..779Y} through which we shoot
light rays to obtain shear maps \citep{2001MNRAS.327..169H} using a
redshift distribution corresponding to the mean $n(z)$ estimated for
the observations. The underlying model is a flat WMAP3-cosmology
($\Omegam=0.27, \Omegab=0.044, \ns=0.95, h=0.71, \sigma_8=0.77$).

An important issue are the correlations between CFHTLS-Wide pointings
due to large-scale structure. Unfortunately, we cannot simulate the
exact survey geometry of the CFHTLS-Wide T0003 survey which extends up
to 8 degrees angular separation, since our simulations are only
$3^{\circ}\times 3^{\circ}$ in size. We therefore use several
(independent) ray-tracing simulations to cover the observed fields;
three for W1, one for W2 and four for W3. This results in
less-correlated fields and a smaller cosmic variance compared to the
observations. We therefore expect the simulations to slightly
underestimate the observed field-to-field variance. The galaxies on
each simulated pointing are distributed homogeneously, we do not take
into account masking. We do however simulate the varying number
density between fields by applying the observed number densities to
the simulations.

\begin{figure}[!tb]

  \resizebox{\hsize}{!}{
    \includegraphics{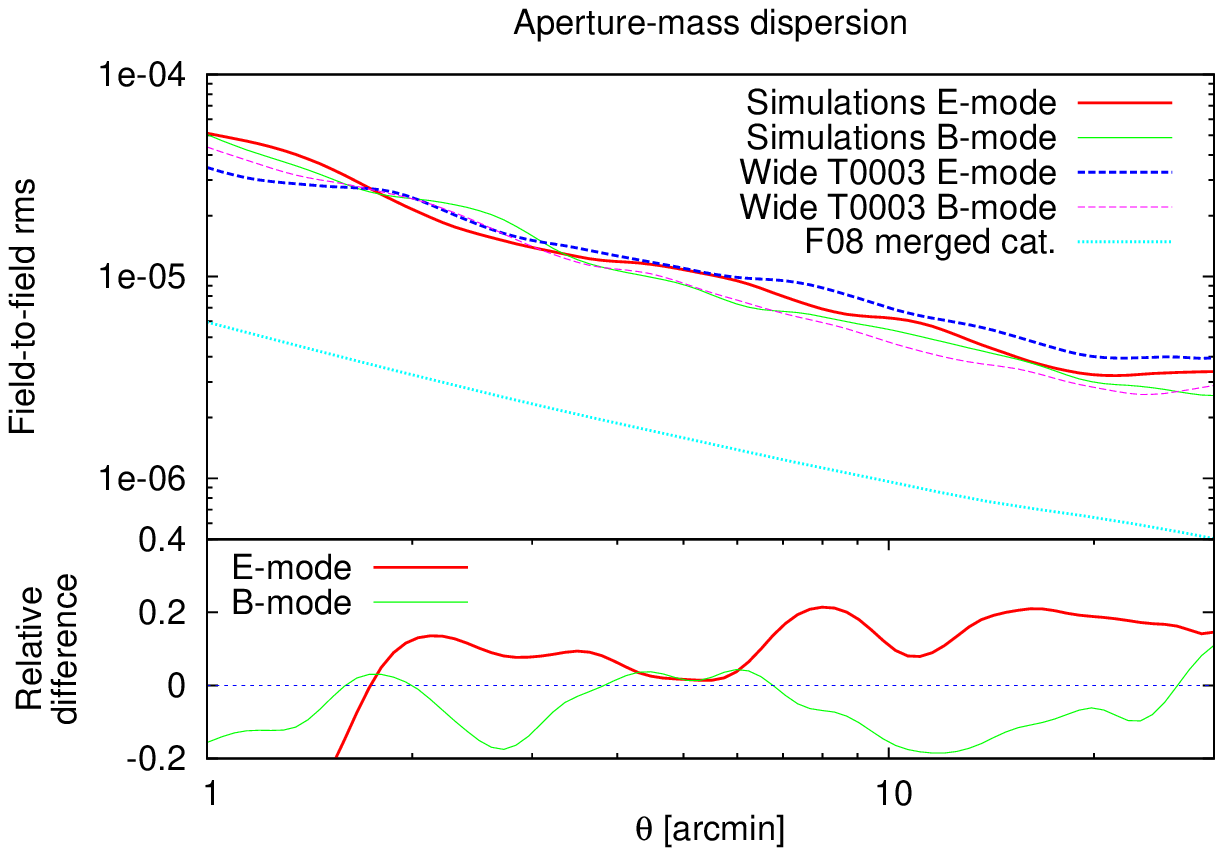}
  }

  \vspace*{1em}

  \resizebox{\hsize}{!}{
    \includegraphics{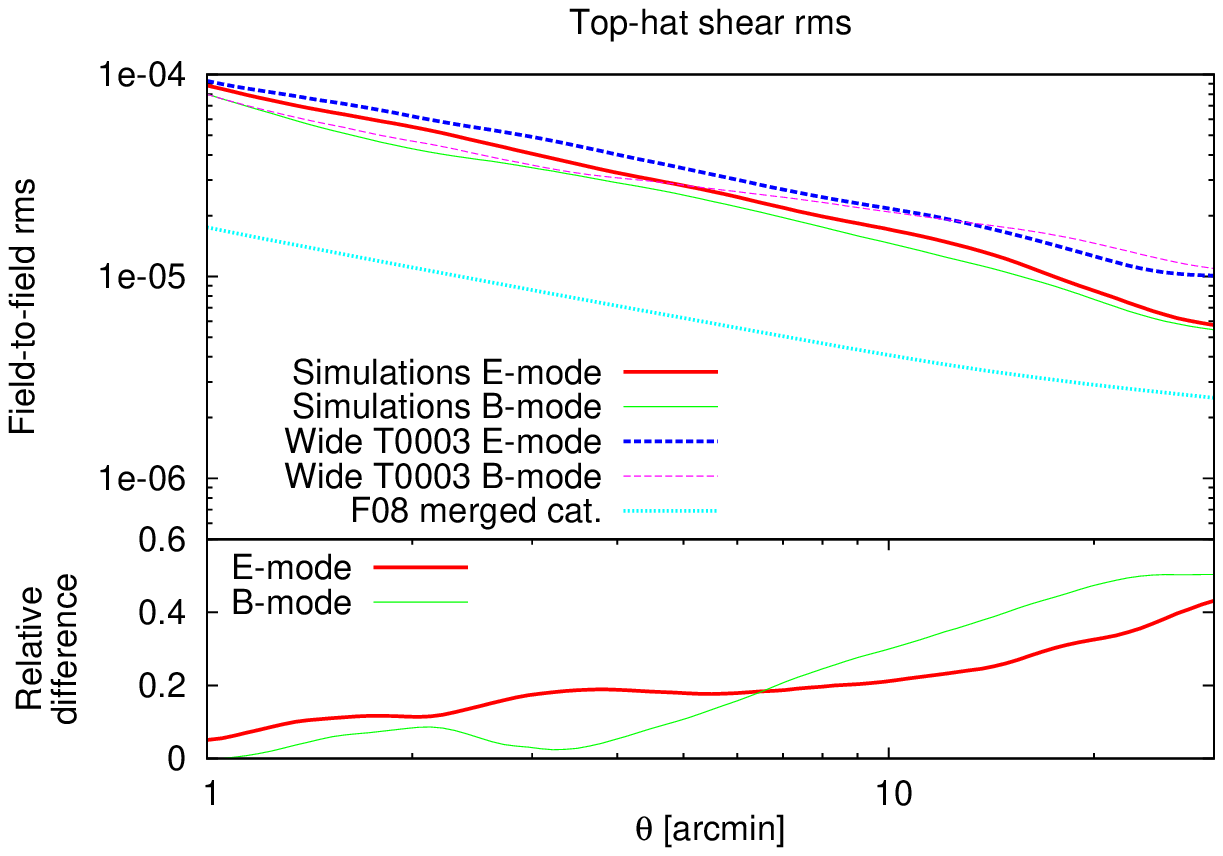}
  }

  \caption{Variance of $\langle M_{\rm ap}^2 \rangle$ from the
    numerical simulations (solid lines) and CFHTLS data (dashed). The
    rms used for the merged catalogue \citep[from][]{FSHK08}, which we
    use for the cosmological constraints, is the dotted line.}
  \label{fig:map2varcomp}
\end{figure}

From Fig.~\ref{fig:map2varcomp} we see that the observed E-mode
variance of $\langle M_{\rm ap}^2 \rangle$ is 10--20\% higher than the
predicted value, on scales between 1 and 30 arc minutes. The B-mode
variance is in very good agreement, with a tendency to be slightly
lower than predicted. From the fact that the B-mode variance is of the
same order as the E-mode, one sees that the considered angular scales
are shape-noise and not cosmic-variance dominated. The power-law shape
of the variance is another indication for this. 
The low-level oscillations on scales on the order of the
$\langle M_{\rm ap}^2 \rangle$-correlation length are probably due to noise.

The observed top-hat field-to-field variance is substantially greater
than the one from simulations, in particular at large angular scales. This
might be due to its higher sensitivity to residuals on large scales
compared to the aperture-mass. Note however that because of the high
correlations between different angular scales for top-hat, the
significance of any discrepancy is hard to interpret. Due to the small
number of simulations we cannot attempt an error estimate on the
field-to-field variance.

Note that the variance for the merged catalogue used in F08, and this
work for the cosmological constraints, is roughly a factor of four
less than the one from individual fields.

The greater observed field-to-field fluctuations could be a sign for
uncorrected residual systematics. We have not included in the
simulations the varying mean redshift due to the different numbers of
exposures for each field in the final stacks. The variance in limiting
magnitude between fields is 0.25. With the empirical law between
limiting magnitude and mean redshift from \citet{2006APh....26...91V}
this translates into a $z$-variance of 0.03. Using the approximate
relation from linear theory $\langle M_{\rm ap}^2 \rangle \propto
z^{1.5}$ we find an additional expected variance of about 4.5\%. We
conclude that the observed E-mode field-to-field variance is higher
than predicted by at most 5\%-15\%. It is difficult to assess the
influence of this additional error on the complete galaxy catalogue
used here. First, we do not know how scales larger than 30 arc minutes
are affected. Second, the shear correlation used to constrain
cosmology is calculated with many more galaxy pairs than in this
field-to-field analysis, with a large number of pairs stemming from
different MegaCam pointings.

\subsubsection{Systematics in the redshift-scaling of the shear
  signal}
\label{sec:systematics-redshift}

Preliminary analysis of weak lensing tomography of the CFHTLS-Wide
shows that beyond a redshift of about unity, the cosmic shear signal
does not increase as expected, but instead is systematically
underestimated \citep{FuPhD}. For this analysis we use the fourth
CFHTLS data release (T0004) which provides five-band photometry on 35
square degrees. Photometric redshifts for each galaxy have been
obtained using the template-fitting code \texttt{Le Phare}
\citep{CIK09}. The quality of the new photo-$z$'s is compatible to
those used in F08 \citep[T0003 data release,][]{2006A&A...457..841I}.
An improvment has been obtained due to additional and larger
spectroscopic samples. We take a sub-set of T0004 consisting of a
contiguous area of 19 square degrees. The measured galaxy
shapes are the same as in F08.

As an illustration, we show the top-hat shear variance measured at 5
and 25 arc minutes for various redshift bins, see Table
\ref{tab:tomo}. The measured values, corresponding to 19 square
degrees, are compared to predictions using a flat $\Lambda$CDM model
with $\Omegam = 0.25$ and $\sigma_8 = 0.8$. For the highest bin with
redshifts above 0.95, the measured values are clearly inconsistent
with the predictions.

\begin{table}[th!]
  \caption{Shear top-hat variance at 5' and 25' and for five redshift
    ranges. The error bars are Poisson noise.}
  \label{tab:tomo}
  \begin{tabular}{l|ll|ll}
     & \multicolumn{2}{c}{$\langle |\gamma^2(5\arcmin)| \rangle/10^5$} &
     \multicolumn{2}{|c}{$\langle |\gamma^2(25\arcmin)| \rangle/10^5$}
     \\ \hline
    $z$-range & CFHTLS & Prediction & CFHTLS & Prediction \\ \hline
    $0.0\phantom{0} \ldots 0.65$  & $-0.3 \pm 2.0$ & 0.95 & $0.2 \pm 0.5$  & 0.29 \\
    $0.0\phantom{0} \ldots 2.5$   & $\phantom{-}1.9 \pm 0.9$  & 2.38 & $0.7 \pm 0.2$  & 0.73 \\
    $0.65 \ldots 0.95$ & $\phantom{-}2.9 \pm 2.2$  & 4.12 & $1.1 \pm 0.5$  & 1.27 \\
    $0.65 \ldots 2.5$  & $\phantom{-}3.3 \pm 1.4$  & 4.77 & $0.9 \pm 0.3$  & 1.48 \\
    $0.95 \ldots 2.5$  & $-1.7 \pm 3.7$ & 7.46 & $0.04 \pm 0.9$ & 2.33 \\
  \end{tabular}
\end{table}

We have to assume that this anomalous redshift-scaling is also present
in the shear catalogue used in this paper for the cosmology
constraints. The redshifts for this preliminary analysis are taken
from the CFHTLS Wide and are therefore slightly less accurate
than the T0003-Deep ones, used to infer the F08 redshift distribution.
However, the former do not suffer from cosmic variance. The
calibration with spectroscopic redshifts assures their reliability for
$i_{AB} \le 24$. The depth of the F08 shear galaxies is only half a
magnitude higher, and those faint galaxies are down-weighted in the weak
lensing analysis. We therefore assume that the problem is mainly due to
the shape measurement and not to the photometric redshifts. One reason for
this could be a shear calibration bias which depends on galaxy
properties that are a function of redshift. A bias in the shape
measurement which is a function of galaxy size, magnitude or galaxy
signal-to-noise will affect low- and high-z galaxies differently.

For this work we choose a simple toy model to parametrise this
potential underestimation of the redshift-scaling of the shear signal,
see \cite{2007JCAP...11..008L} for a similar approach. This
decrease of the lensing signal corresponds to an effective lensing
efficiency which is lower than expected at high redshifts. To model
this we multiply the redshift distribution $n(z)$ with a constant
$c_0>0$, for redshifts $z>z_0$. We chose $z_0$ to be 1.0. $c_0=1$ is
the unbiased case with no degradation of the shear signal. $c_0=0$ is
the pessimistic case where there is no shear signal for $z>z_0$. We
allow values greater than unity corresponding to an overestimation of
the shear signal. Note that we use the original, unaltered redshift
distribution when fitting the $n(z)$ histogram (eq.~\ref{chi2nofz}).

\begin{figure}[!tb]
  
  \begin{center}
  \resizebox{\hsize}{!}{
    \hspace*{3em}\includegraphics{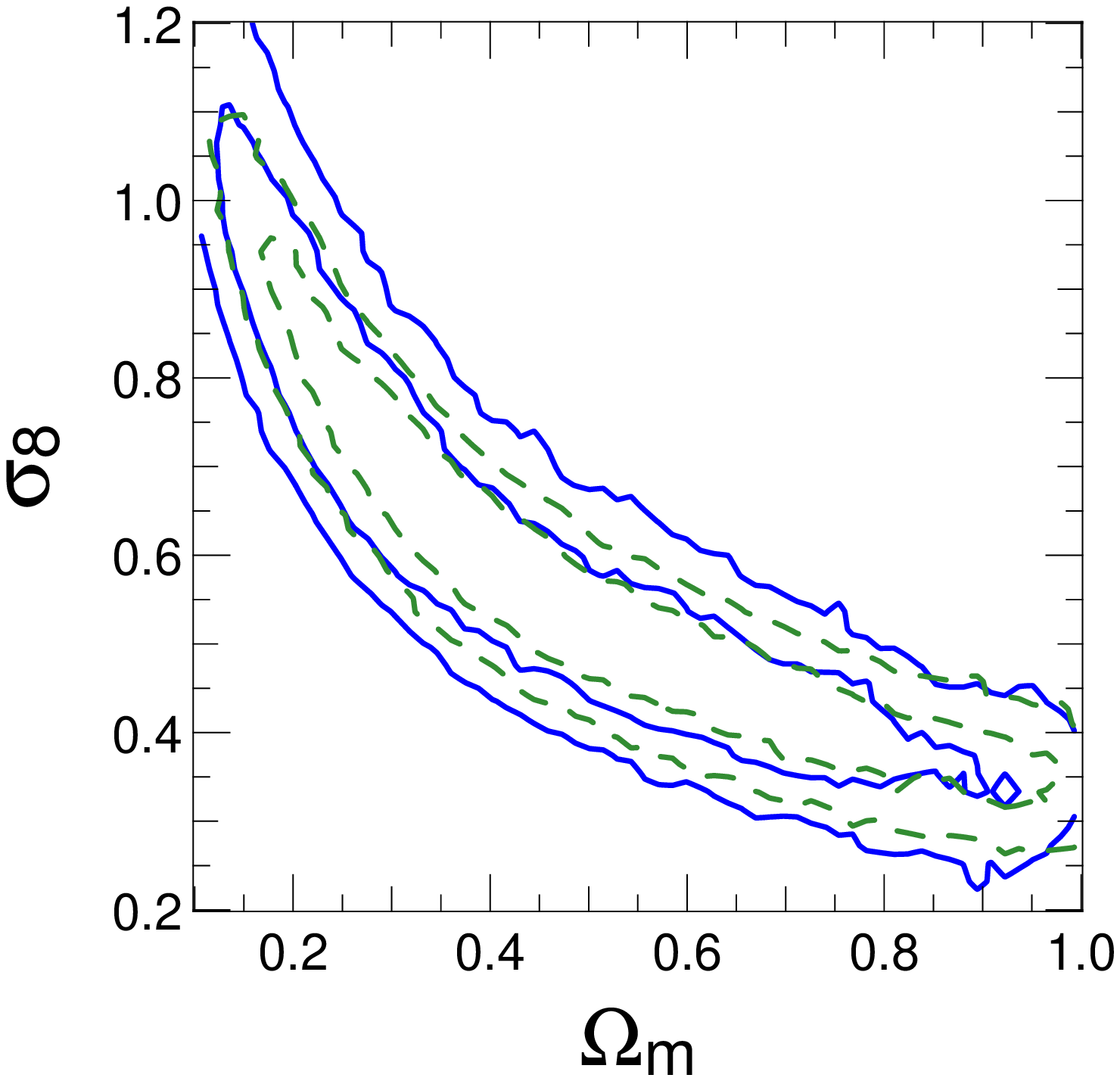}
    \hspace*{8em}\includegraphics{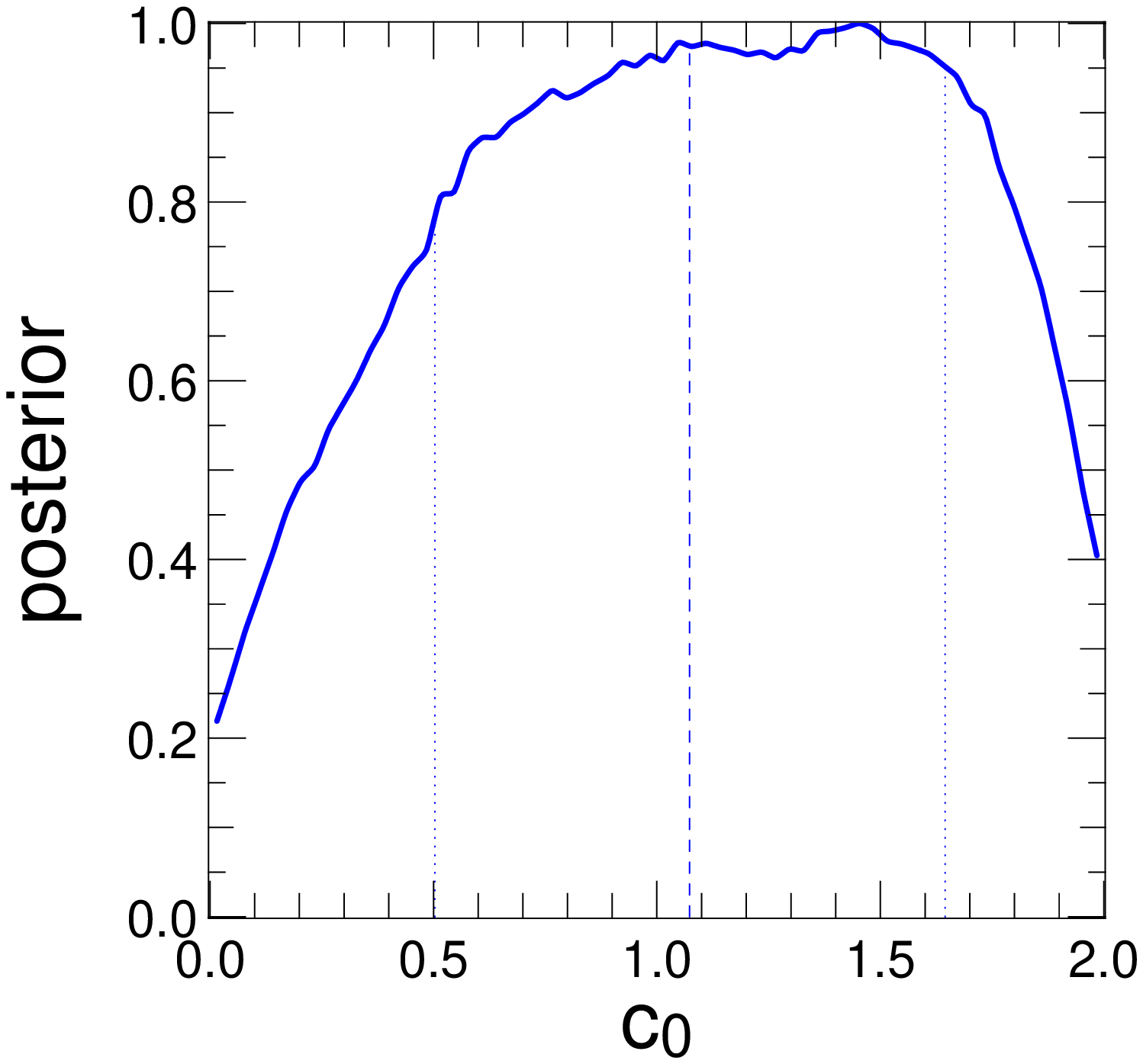}
  }
  \end{center}
  
  \caption{\emph{Left panel:} 2d marginals (68\% and 95\%) for
    $\Omegam$ and $\sigma_8$ for the two cases of including the
    high-$z$ calibration bias (solid lines) and ignoring it (dashed
    curves), respectively. \emph{Right panel:} 1d marginal likelihood
    for $c_0$. The vertical lines indicate mean (dashed) and 68\%
    confidence intervals (dotted).
  }
  \label{fig:Om_s8_lens_comp_c0}
\end{figure}

\begin{figure}[!tb]

  \begin{center}
    \resizebox{\hsize}{!}{
      \hspace*{5em}\includegraphics{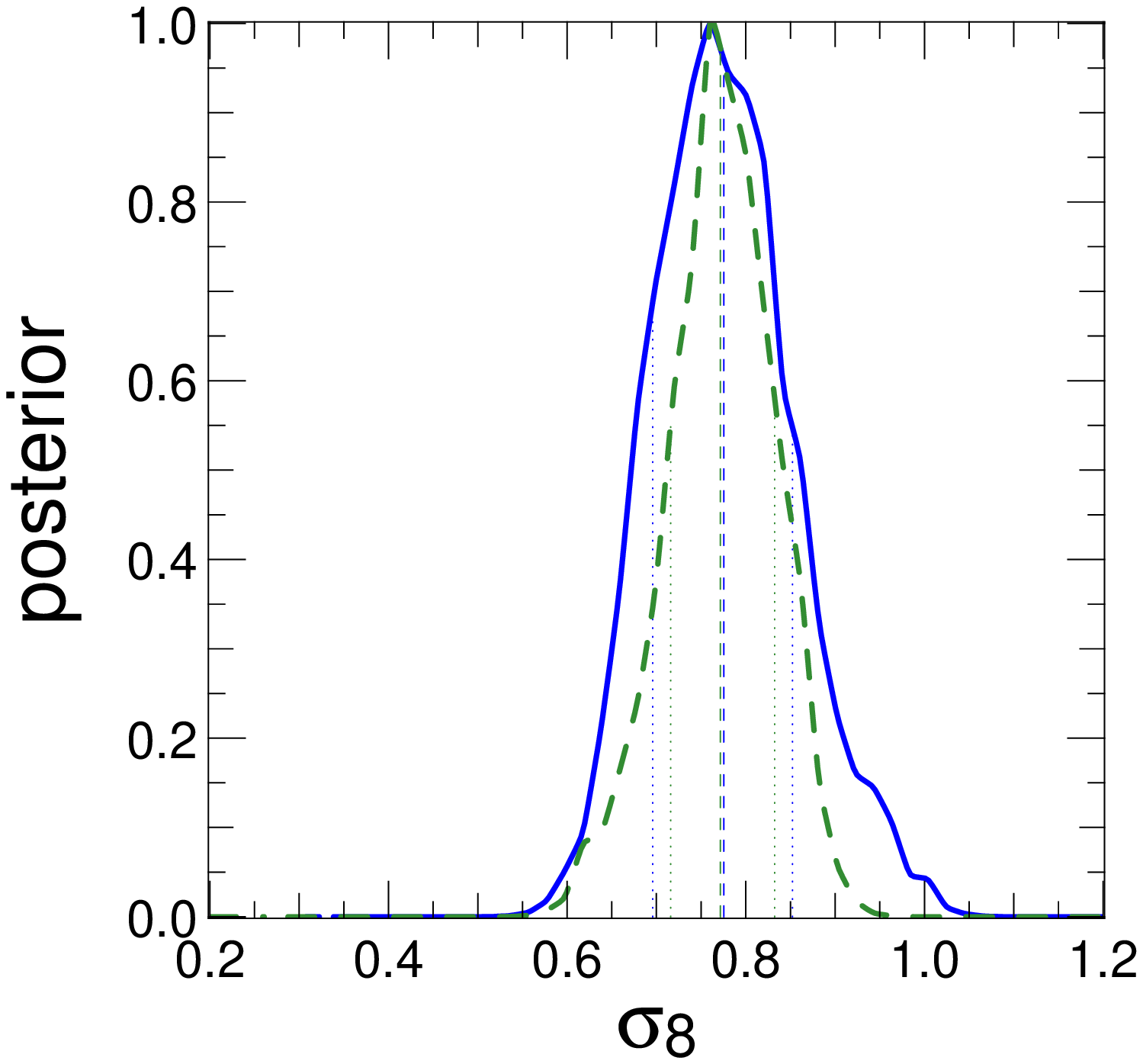}
      \hspace*{3em}\includegraphics{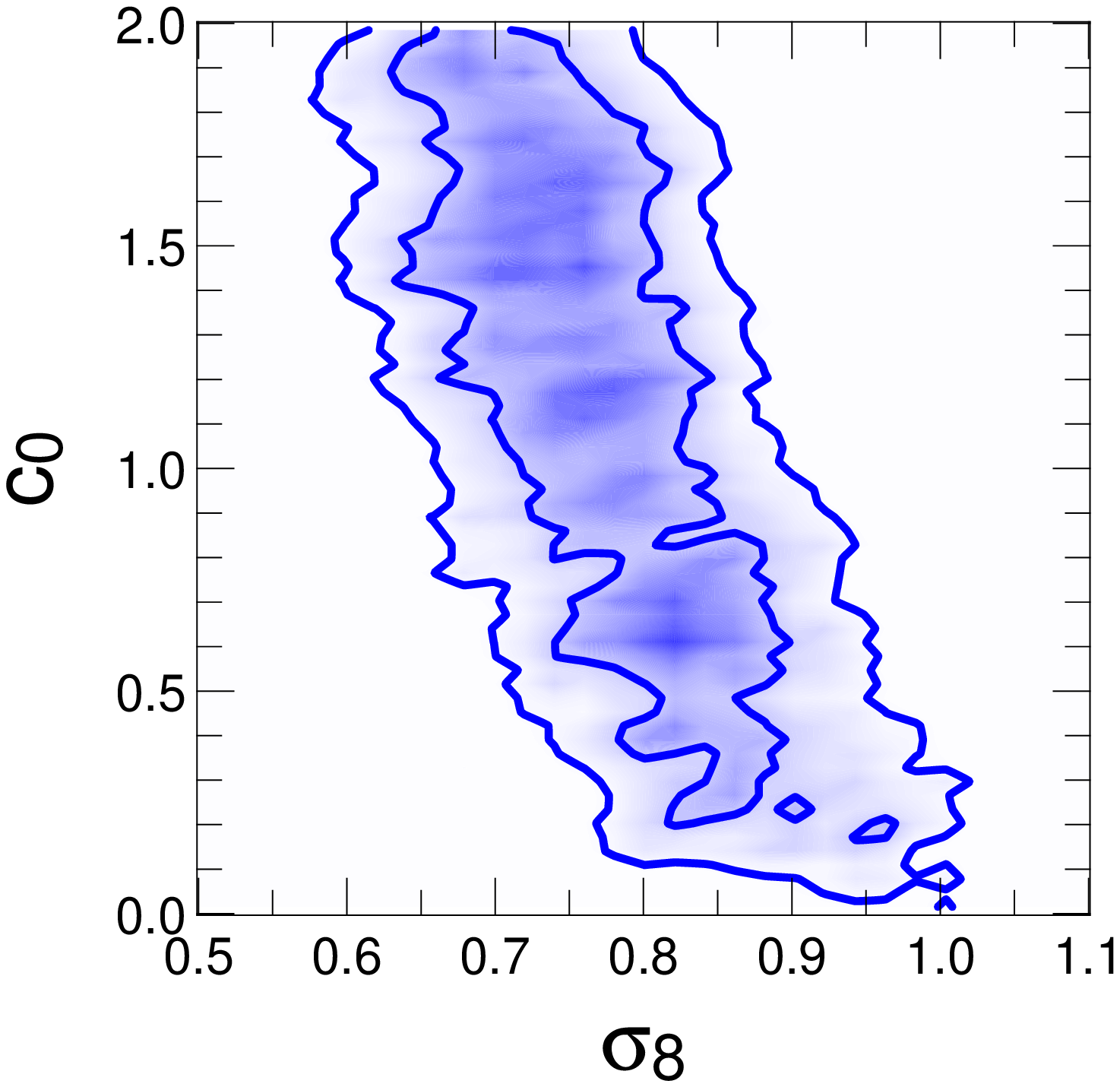}
    }
  \end{center}

  \caption{
    \emph{Left panel:} 1d marginal for $\sigma_8$ for $\Omegam =
    0.25$, the vertical lines are as in
    Fig.~\ref{fig:Om_s8_lens_comp_c0}.
    \emph{Right panel:} Correlation between $c_0$ and
    $\sigma_8$ (68\% and 95\% confidence levels) for $\Omegam =
    0.25$. The vertical lines indicate mean (dashed) and 68\%
    confidence intervals (dotted).
  }
  \label{fig:s8_c0}
\end{figure}

Using a model with fixed $\Omegab$ and $\ns$ and marginalising over
$h, \sigma_8$ and redshift parameters we measure a value of $c_0$
consistent with unity but with large error margins, $c_0 = 1.1 \pm 0.6 \,
(68\%)$ (see also Fig.~\ref{fig:Om_s8_lens_comp_c0}, right panel). At
95\% confidence nearly the whole range of $c_0$ is permitted. Ignoring
the high-$z$ calibration error leads to smaller error bars for other
parameters, in particular $\sigma_8$ (left panel of
Fig.~\ref{fig:Om_s8_lens_comp_c0}). For a fixed $\Omegam = 0.25$ we
respectively get $\sigma_8 = 0.78^{+0.08}_{-0.07}$ and $\sigma_8 = 0.77 \pm
0.05$ with and without modelling the systematic error. Because we allow the
calibration to be smaller and greater than unity there is no
significant bias on $\sigma_8$. If we restrict $c_0$ to values smaller
or equal 1, thus using the information about the underestimation of
the shear signal as prior, we obtain a best-fit $\sigma_8$ of 0.83
($\Omegam = 0.25$).  Thus, $\sigma_8$ might be underestimated by $8\%$
which is about one sigma, provided our simple model is correct. From
Fig.~\ref{fig:s8_c0} (left panel) we see the potential shear measurement
systematics shifts the likelihood and introduces a tail for high
$\sigma_8$.

We repeat the above analysis with $z_0 = 0.8$ and 1.2 and find only
small shifts $\Delta \sigma_8$ for the power-spectrum
normalisation. Writing $\Delta \sigma_8 = \lambda \, (z_0 - 1)$, we
find $\lambda = -0.03$ if $c_0$ is varied between 0 and 2, and
$\lambda = -0.09$ for $c_0<1$, corresponding in both cases to a 2\%
change for $\sigma_8$ in the considered range for $z_0$.

Interestingly, we see no correlation between $c_0$ and any other individual
parameter when the whole posterior is considered. However, $c_0$ is of
course correlated with the combination of $\Omegam$ and $\sigma_8$
which determines the shear amplitude. When fixing
$\Omegam$ we see a strong correlation with $\sigma_8$, as expected
(Fig.~\ref{fig:s8_c0}, right panel).

To summarise the results on weak lensing systematics, we note that
this study is far from complete. Other tests focusing on the PSF
correction have been made in F08, and extensive studies going into
much more detail as presented here will be published soon (van
Waerbeke et al.~in prep.).

\subsection{SNLS supernovae Ia}
\label{sec:SN_like}

The supernova data set corresponds to the first data release of
the SNLS, see A06. We use 105 supernovae in
total, 71 of which were observed with the CFHTLS. Their redshift range
is between 0.25 and 1.
The remaining 44
objects are nearby supernovae, with $0.015<z<0.13$. In the next
section (\ref{sec:SN_like_standard}) we describe the ingredients of
the standard likelihood analysis, ignoring systematics, which is
similar to A06 with slight modifications. Section
\ref{sec:SN_like_sys} introduces the photometric calibration
parameters and defines the updated likelihood function including
systematics.

\subsubsection{Standard likelihood analysis}
\label{sec:SN_like_standard}

We use the results from the SNIa light-curve fits of A06. For each
supernova the fit yields the rest-frame $B$-band magnitude $m_B^*$,
the shape or stretch parameter $s$, and the colour $c$. These
quantities are combined in the estimator of the distance modulus for the $i^{\rm th}$
object as follows:
\begin{equation}
  \mu_{B, i} = m_{B, i}^* - \bar M + \alpha (s_i-1) - \beta c_i.
  \label{mu0_SN}
\end{equation}
The universal absolute SNIa magnitude is $\bar M$; $\alpha$ and
$\beta$ are universal linear response factors to stretch and colour,
respectively.

The error needed for the likelihood includes: (1) The measured
uncertainties of the light-curve parameters $m_{B, i}^*, s_i, c_i$ and
their correlations. The corresponding error contribution to $\mu_{B,
  i}$ can be written as $\sigma^2(\mu_{B, i}) = \vec \theta_2^{\, \rm
  t} W_2^{} \vec \theta_2^{}$ with the parameter vector $\vec \theta_2
= (1, \alpha, \beta)$ and the covariance $W_2$ of the data vector
$(m_{B,i}^*, s_i, c_i)$. (2) An additional redshift uncertainty
$\sigma_{{\rm pv}, i} = 5/\ln 10 \cdot v_{\rm p}/(c \, z_i)$ due to
peculiar velocities corresponding to $v_{\rm p} = 300$ km
s$^{-1}$. (3) An intrinsic scatter in absolute magnitude of
$\sigma_{\rm int} = 0.13$.

Those three errors are added in quadrature to the log-likelihood
expression, which is
\begin{align}
  \chi^2_{\rm sn}(\vec p) = & \sum_i \frac{\left[
      \mu_{B,i}(\vec p) - 5 \log_{10}\left(\frac{d_{\rm L}(z_i, \vec p)}{10\,{\rm pc}}
      \right)
      \right]^2}{\sigma^2(\mu_{B, i}) +  \sigma_{{\rm pv}, i}^2 +
    \sigma_{\rm int}^2},
  \label{chi20_SN}
\end{align}
The linear dependence on the Hubble constant of the luminosity
distance $d_{\rm L}$ is taken out as an additional summand in
eq.~(\ref{chi20_SN}) and integrated into the absolute luminosity. From
now on we will use the parameter $M = \bar M - 5 \log_{10} h_{70}$.

The error term from the light-curve fit, $\sigma^2(\mu_{B,i})$,
  depends on the stretch and colour responsitivies $\alpha$ and
  $\beta$. A06 kept those parameters fixed during the
  $\chi^2$-minimisation and updated them iteratively in consecutive
  minimisation runs.
  This leads to a bias in those parameters of 7\% - 15\%.
  Instead, we include the dependence of $\alpha$ and $\beta$ in the
  denominator in eq.~(\ref{chi20_SN}) to obtain unbiased best-fit
  results for those parameters which are therefore larger than those
  cited in A06, see Table \ref{tab:S_sys}.

We do not take into account the term involving the covariance
determinant in the log-likelihood.
Even though in this case this term is not a constant because of the
parameter-dependent variance, we
verified that the effect on cosmological parameters is very small.
Moreover, it is in accordance to most other
SNIa analyses \cite[e.g.][]{2006A&A...447...31A,kowalski-2008}.

\subsubsection{Extended analysis with systematics}
\label{sec:SN_like_sys}

We extend the analysis of A06 by including the response of the
distance modulus to a photometric zero-point shift in each of the
seven filters $(g, r, i, z, U, B, V)$ and in the Vega $(B-R)$ colour in
the Landolt system. The reference photometric zero-points are those
used in A06, which were estimated without using the Hubble diagram. If
$\vec \theta_1$ denotes the vector of those eight zero-point magnitude
shifts, the linear response $k_{i \alpha}$ is the change in distance
modulus for the $i^{\rm th}$ supernova for a small change of the
$\alpha^{\rm th}$ zero-point,
\begin{equation}
  k_{i \alpha} = \frac{\Delta \mu_{B, i}}{\Delta \theta_{1 \alpha}};
  \;\;\;\; \alpha = 0 \ldots 7.
  \label{SN_ki}
\end{equation}
By changing the zero-points and redoing the light-curve fit, we
obtain the values of $k_{i\alpha}$ for each supernova.  We include this
change to the distance modulus linearly in the model, which results in an
additional term $\vec k_i^{\rm t} \cdot \vec \theta_1$ in the
likelihood.

We assume that the zero-point magnitude shift parameters $\vec
\theta_1$ are uncorrelated variables, since the data in different
optical bands have been reduced independently. The noise is the result
from a number of different source and can therefore be well
approximated to be Gaussian. By definition they mean of $\vec
\theta_1$ is zero. The rms is taken to be $0\,\fm01$ for all filters
except for $z$ ($\theta_{13}$) where we assume the rms to be
$0\,\fm03$. The numerical values of those uncertainties are taken from
A06 (Sect.~4.1) and correspond to the limits of reproducibility of the
photometric calibration. The corresponding (diagonal) covariance
matrix is $W_1$.

This prior information is multiplied to
the SNIa-likelihood in the form of a multi-variate Gaussian
likelihood. With the additional term, the log-likelihood expression
corresponding to the extended analysis is then
\begin{align}
  \chi^2_{\rm sn+sys}(\vec p) = & \, \vec \theta_1^{\rm t}
  W_1^{-1} \vec \theta_1^{} \nonumber \\
  & + \sum_i \frac{\left[ \mu_{B,i}(\vec p)
      + \vec k_i^{\rm t} \cdot \vec \theta_1 - 5
      \log_{10}\left(\frac{d_{\rm L}(z_i, \vec p))}{10{\rm pc}} \right)
      \right]^2}{\sigma^2(\mu_{B, i}) + \sigma_{{\rm pv}, i}^2 +
    \sigma_{\rm int}^2}.
    \label{chi2E_SN}
\end{align}
Now, the parameter vector $\vec p$ contains the zero-point parameters
$\vec \theta_1$. The corresponding optical bands are indicated in
Table \ref{tab:syspar}.

\begin{table}[!th]

  \caption{List of SNIa systematic parameters and their symbols.}
  \label{tab:syspar}

  \begin{tabular}{ccl}\hline\hline
    $M$     & & \\ 
    $\alpha$ & $\theta_{22}$   & Light-curve parameters \\
    $\beta$ & $\theta_{23}$   & \\ \hline
    $\Delta g$ & $\theta_{10}$ & \\
    $\Delta r$ & $\theta_{11}$ & \\
    $\Delta i$ & $\theta_{12}$ & \\ 
    $\Delta z$ & $\theta_{13}$ & Zero-point shifts \\
    $\Delta U$ & $\theta_{14}$ & \\ 
    $\Delta B$ &   $\theta_{15}$  & \\
    $\Delta V$ & $\theta_{16}$ & \\
    $\Delta(B-R)_{\rm Vega}$   & $\theta_{17}$ \\ \hline
  \end{tabular}

\end{table}

\subsubsection{Systematics for SNIa}

Unlike systematics in weak lensing shape measurements which are
difficult to model, the observation-related systematics for SN are
more easily parametrised. In the next section we take into account
errors in the estimated distance modulus due to uncertainties in the
photometric calibration. The results of the MCMC analysis for the SNIa
internal parameters are given in Table \ref{tab:S_sys}.

\subsubsection{Bias due to systematics}
\label{sec:bias_sys}

\begin{figure*}[!tb]
  \sidecaption  
    \resizebox{0.7\hsize}{!}{
      \includegraphics{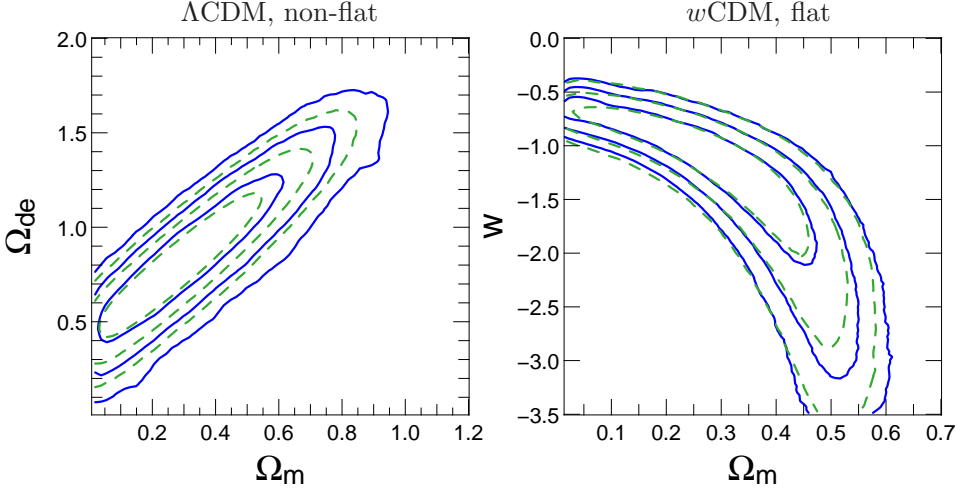}
    }
  \caption{Confidence contours (68\%, 95.5\%, 99.7\%) for full
    treatment of errors due to photometry zero-points (blue solid
    lines) and ignoring those errors (green dashed curves). These two
    cases correspond to the two cases in Table \ref{tab:zeropt}. The
    left panel corresponds to $w=-1$, the right panel is for a flat
    $w$CDM Universe. The constraints are obtained using SNIa alone.}
  \label{fig:zeropt2d}
\end{figure*}

We compare the case of ignoring the systematic errors, using
eq.~(\ref{chi20_SN}) as the log-likelihood, with the case of fully
taking into account the systematics according to the log-likelihood
(\ref{chi2E_SN}).  As can be seen in Fig.~\ref{fig:zeropt2d}, ignoring
the zero-point errors leads to an asymmetric decrease of the error
bars. The constraints get tighter, mainly along the direction of
constant luminosity distance, which is the parameter-degeneracy
direction.  The error bars decrease by about 10\%, see Table
\ref{tab:zeropt}. The bias on parameter means is small, between 10\%
and 20\% of the statistical uncertainty. The bias on the intrinsic
SNIa parameters ($M, \alpha, \beta$) is even smaller, not more than a
few percent of the statistical uncertainty.
For a fixed $\Omegam = 0.25$, the absolute biases on both the eos
parameter (for $\Omegade = 1 - \Omegam$) and the dark-energy density
(for $w=-1$) are smaller than for the marginalised case, but remain to
be about a tenth of the statistical error.

\begin{table}[b!]

  \caption{Parameter means and 68\%-confidence intervals when,
    respectively, ignoring
    and taking into account the systematic errors in the form of
    photometric zero-point fluctuations, see
    Sect.~\ref{sec:bias_sys}.}

  \label{tab:zeropt}

  \renewcommand{\arraystretch}{1.4}
  \begin{tabular}{c|cc}\hline\hline
    \multicolumn{3}{c}{$w=-1$} \\
               & With systematics & Ignoring systematics \\ \hline
    $\Omegam$  & $0.34^{+0.21}_{-0.18}$
               & $0.30^{+0.19}_{-0.16}$ \\
    $\Omegade$ & $0.87^{+0.30}_{-0.27}$
               & $0.83^{+0.27}_{-0.23}$ \\ 
    $\Omegade(\Omegam=0.25)$ & $0.76^{+0.10}_{-0.12}$
                             & $0.75^{+0.09}_{-0.09}$ \\
    \hline\hline
    \multicolumn{3}{c}{flat Universe} \\
               & With systematics & Ignoring systematics \\ \hline
    $\Omegam$  & $0.32^{+0.11}_{-0.20}$
               & $0.31^{+0.11}_{-0.17}$ \\
    $w$        & $-1.38^{+0.46}_{-0.91}$
               & $-1.32^{+0.39}_{-0.80}$ \\
    $w(\Omegam=0.25)$ & $-1.00^{+0.12}_{-0.12}$
                      & $-1.01^{+0.09}_{-0.10}$ \\
    \hline \hline
  \end{tabular}
  \renewcommand{\arraystretch}{1}

\end{table}

As mentioned in Sect.~\ref{sec:SN_like}, we obtain unbiased best-fit
values for the stretch and colour response parameters, $\alpha$ and
$\beta$, respectively (see Table \ref{tab:S_sys}). These differ by
about 15\% from the (biased) values given in A06. The absolute
magnitude $M$ is consistent with A06.

The parameters describing the zero-point shifts ($\theta_{10}$ to
$\theta_{17}$) are all consistent with zero (Table
\ref{tab:S_sys}). Except for $\theta_{13} = \Delta z$ they have zero
mean and rms of about $0\,\fm01$. The influence of the derivative
(eq.~\ref{SN_ki}) on the second term of the likelihood
(eq.~\ref{chi2E_SN}) is small in comparison with the first
term. Nevertheless, correlations with other parameters are introduced
as is discussed in the next section. The mean of the $z$-band
zero-point shifts $\theta_{13}$ is negative (although not
significantly so) and its variance smaller than the expected value of
$0\,\fm03$. It is also this parameter which shows the highest
correlation with cosmological parameters.

\begin{table}[th]
  \caption{Mean and 68\% errors for SNIa internal parameters, for a
    flat $w$CDM model. The values of $\theta_{1i}, i=0\ldots 7$ are in units of
    0.01 magnitudes.}
  \label{tab:S_sys}
  \renewcommand{\arraystretch}{1.5}
  \begin{center}\begin{tabular}{|l|l|}\hline
\rule[-3mm]{0em}{8mm}Parameter	 & Best-fit-value	\\ \hline\hline
$-M$	 & $19.337^{+0.036}_{-0.041}$	\\ \hline
$\alpha$	 & $1.62^{+0.14}_{-0.15}$	\\ \hline
$-\beta$	 & $-1.80^{+0.17}_{-0.16}$	\\ \hline
$\theta_{10}=\Delta{g}$	 & $-0.06^{+0.99}_{-1.00}$	\\ \hline
$\theta_{11}=\Delta{r}$	 & $0.10^{+0.98}_{-0.96}$	\\ \hline
$\theta_{12}=\Delta{i}$	 & $0.09^{+0.95}_{-0.95}$	\\ \hline
$\theta_{13}=\Delta{z}$	 & $-1.0^{+2.5}_{-2.5}$	\\ \hline
$\theta_{14}=\Delta{U}$	 & $-0.05^{+0.98}_{-0.99}$	\\ \hline
$\theta_{15}=\Delta{B}$	 & $-0.04^{+0.99}_{-0.98}$	\\ \hline
$\theta_{16}=\Delta{V}$	 & $-0.06^{+0.98}_{-0.97}$	\\ \hline
$\theta_{17}=\Delta(B-V)_{\rm\,Vega}$	 & $-0.05^{+0.99}_{-1.01}$	\\ \hline
\end{tabular}\end{center}

  \renewcommand{\arraystretch}{1}
\end{table}

\subsubsection{Correlation between systematic errors and cosmological parameters}

\begin{figure}[!tb]

  \begin{center}
    \resizebox{\hsize}{!}{
      \hspace*{5ex}\includegraphics{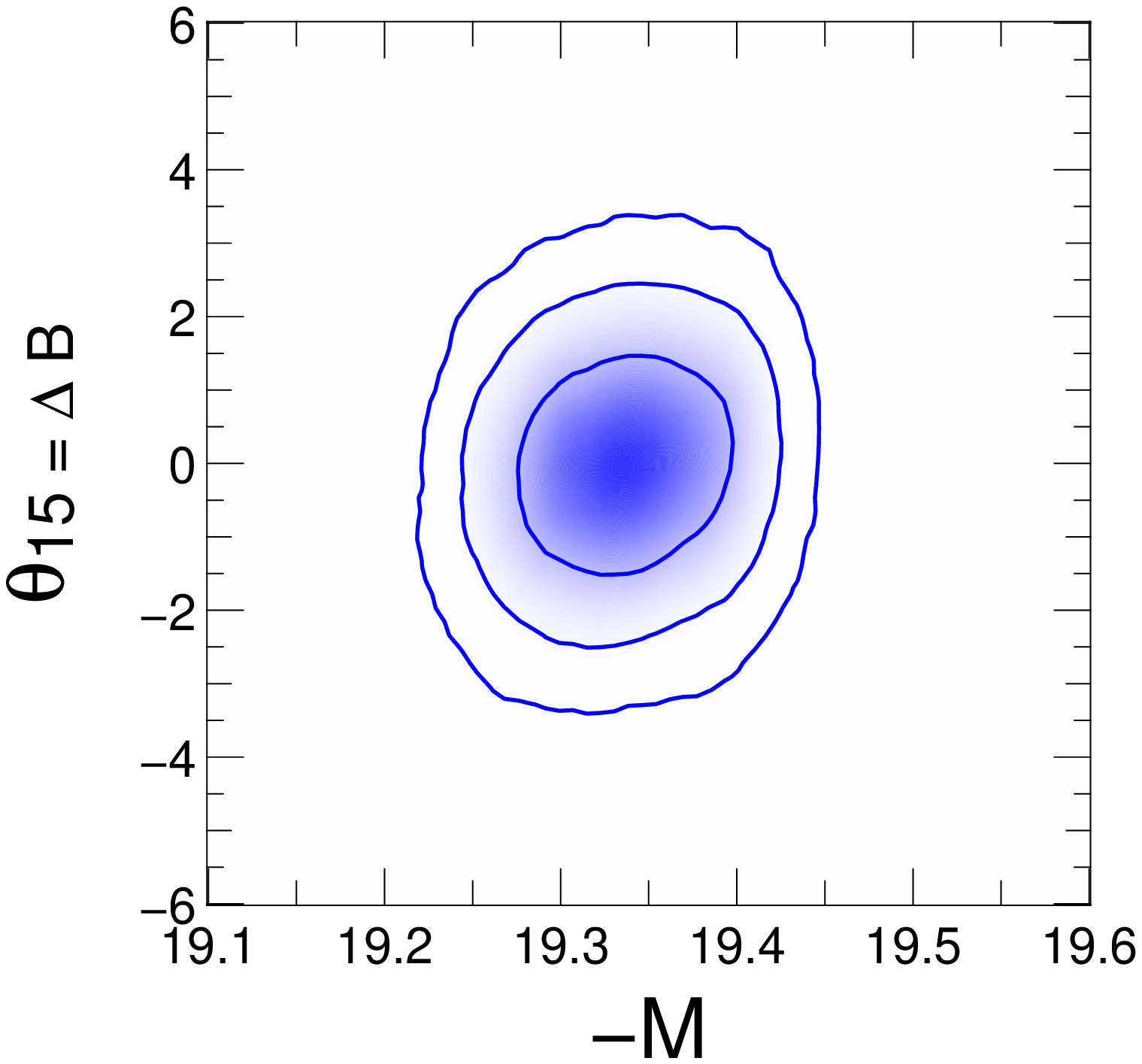}
      \hspace*{5ex}\includegraphics{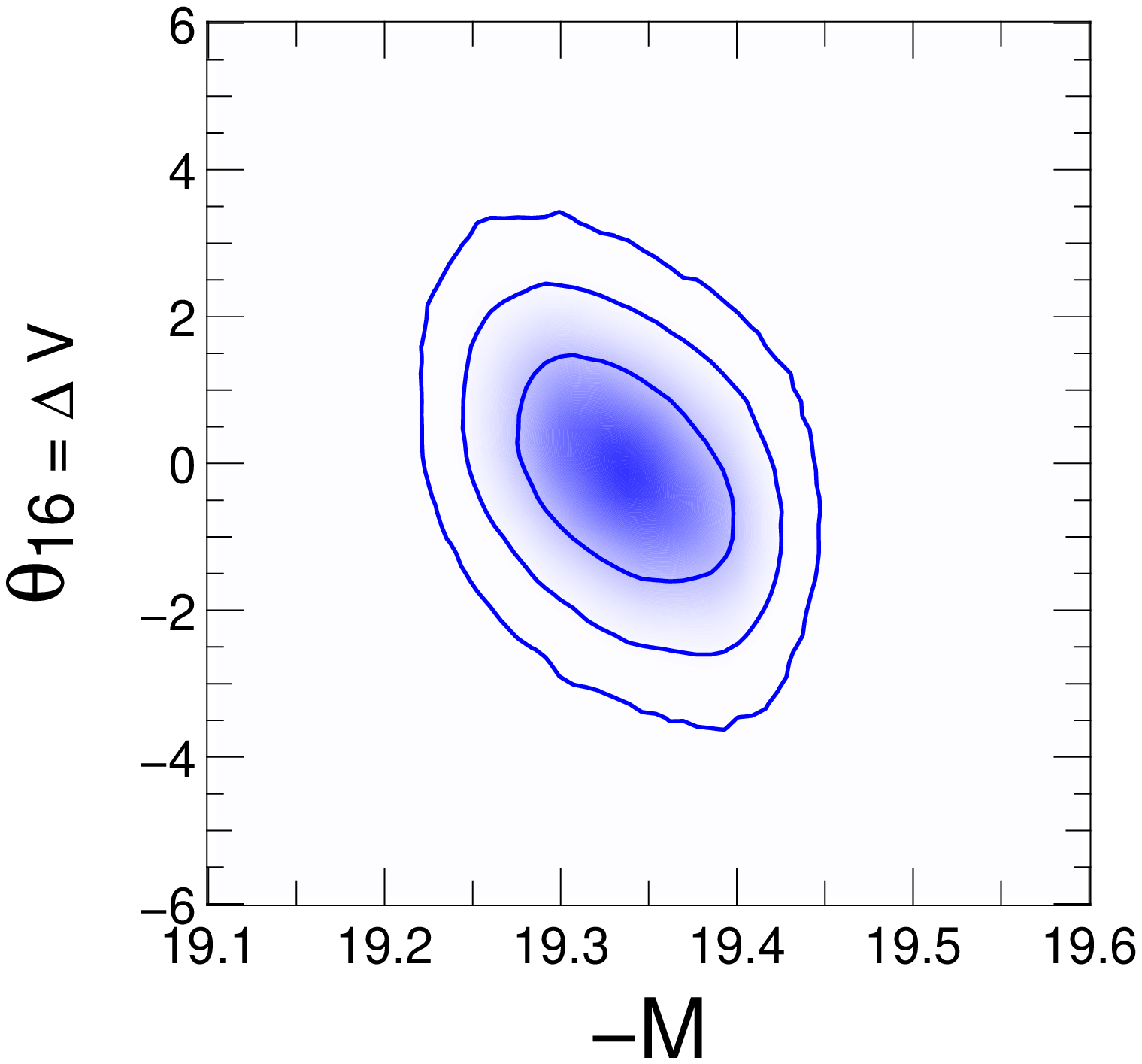}
    }
  \end{center}

  \caption{Correlations between the universal absolute SNIa
    magnitude $M$ and the $B$-band ($\theta_{15}$, left panel) and
    $V$-band ($\theta_{16}$, right panel), respectively. The
    zero-point shift parameters $\theta_{1i}$ are given in units of
    0.01 magnitudes.}
  \label{fig:theta-M}
\end{figure}

In most cases the zero-point shifts $\vec \theta_1$ are
uncorrelated with other parameters. However, some pairs
of one or more nuisance parameters show correlations, most notably, the
absolute SNIa magnitude $M$ is correlated both with the $B$- and
$V$-band uncertainty $\Delta B = \theta_{15}$ and $\Delta V =
\theta_{16}$, respectively, see Fig.~\ref{fig:theta-M}. This
correlation can be explained by looking at the rest-frame colour
parameter, $c= (B-V)_{B {\rm max}} + 0.057$, which contains the $B-V$
colour excess at the time of the $B$-band maximum. Redefining the $B$-
or $V$-band zero-points causes a systematic change in $c$ which is
compensated by a corresponding change in $M$. The slope of the
measured $M-\Delta V$-correlation is about 1.5. The correlation
between $M$ and $\Delta B = \theta_{16}$ is much weaker and has a
smaller slope of -0.3 to -0.4. The correlation between $M$ and the
difference $\Delta B - \Delta V$ is therefore on the order of the
best-fit value of $\beta = 1.80$, which is expected from
eq.~\ref{mu0_SN}. The difference in correlation can be explained by
the fact that the observed magnitudes of each object translate into
different rest-frame bands for the light-curve fit depending on the
object's redshift.

\begin{figure}[!tb]

  \begin{center}
    \resizebox{\hsize}{!}{
      \hspace*{5ex}\includegraphics{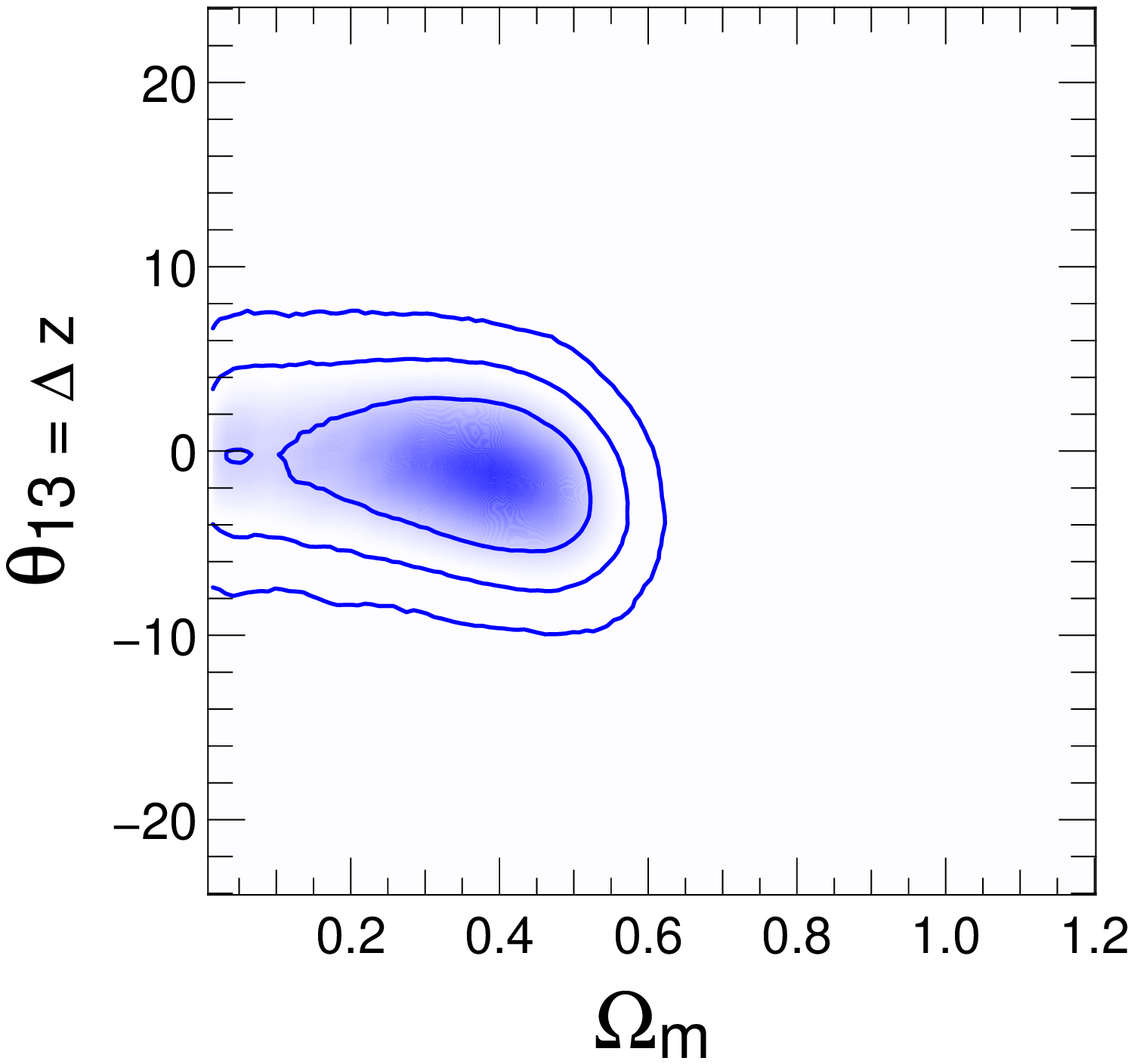}
      \hspace*{5ex}\includegraphics{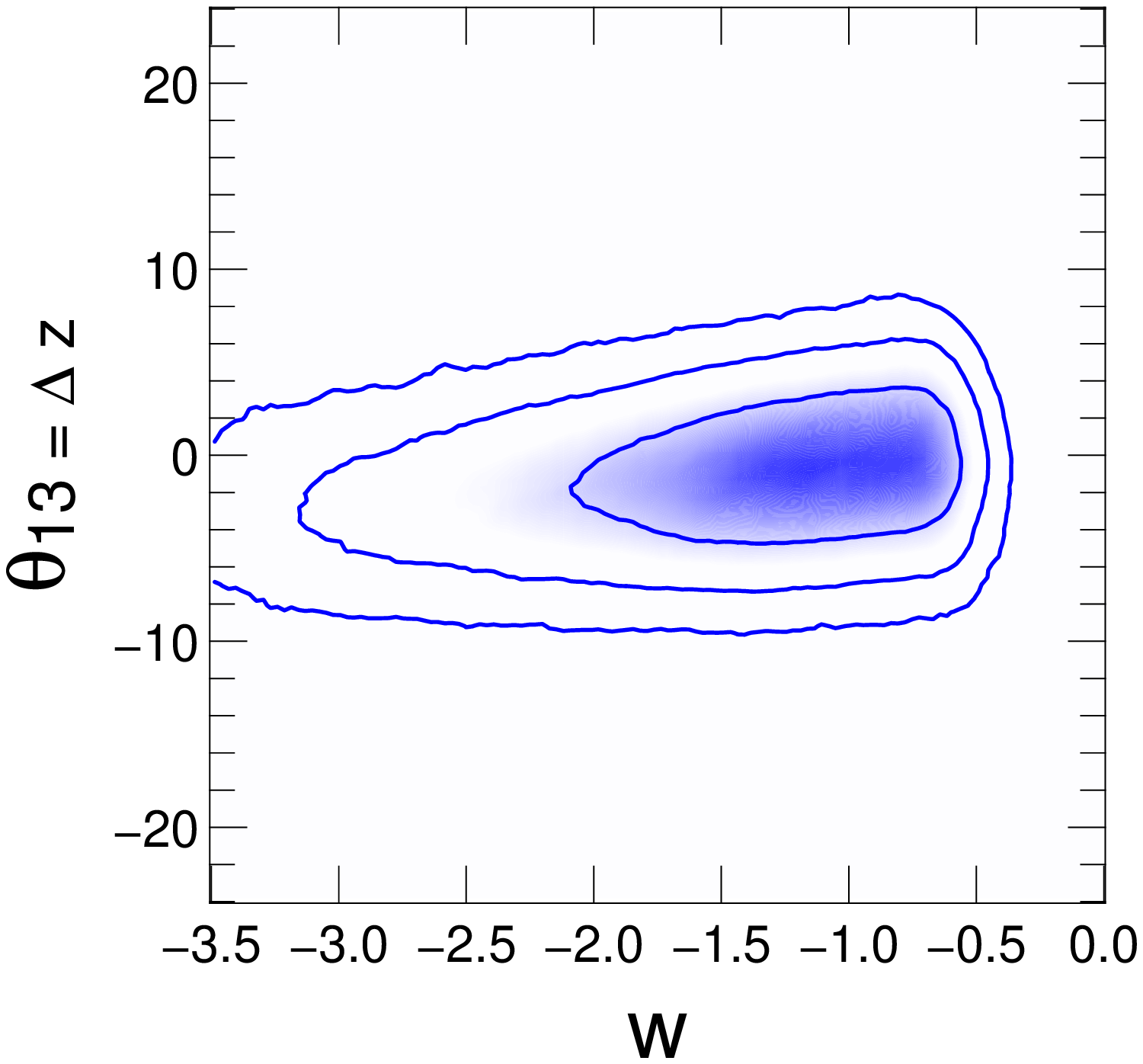}
    }
  \end{center}

  \caption{Correlations between the $z$-band zero-point offset
    ($\theta_{13}$) and $\Omegam$ (left panel) respectively $w$ (right
    panel).}
  \label{fig:theta-Ow}
\end{figure}

Further, the $z$-band zero-point offset ($\theta_{13}$) shows a
correlation with both $\Omegam$ and $w$ (Fig.~\ref{fig:theta-Ow}).
We find $\Delta w = 0.057 \, \Delta \theta_{13}/(0.03 \, {\rm mag})$
and $\Delta \Omegam = -0.0159 \, \Delta \theta_{13}/(0.03 \, {\rm
  mag})$. The latter value can be compared to A06 (Table 5) who varied
the zero-points to infer the influence on cosmological parameters; our
value is of the same order of magnitude as the one from A06 but has
opposite sign.  For a fixed $\Omegam=0.25$ we obtain $\Delta w =
-0.0135 \, \Delta \theta_{13}/(0.03 \, {\rm mag})$, about half of the
value cited in A06 (Table 5).

\subsubsection{Further sources of systematics}

A06 discussed further sources of systematic uncertainties which can
affect cosmological results from SNIa. Some of those uncertainties
could be parametrised and invoked in a joint analysis with cosmology
as presented in this paper. We state a few more examples here and leave a
thorough analysis for future work.

For example, the difference between
measured rest-frame $U$-band magnitude and the one `predicted' from the
light-curve fit using only two bands can be included in the likelihood
function.

In addition, the Malmquist bias can be modelled as a function of
redshift. The inclusion of this bias will be necessary when
redshift-dependent dark-energy models are to be tested. This requires
accurate knowledge of the sample and of selection effects.

Furthermore, weak gravitational lensing of distant SNIa can be modelled
using the dark-matter power spectrum, either from (non-linear)
theoretical prescriptions \citep{2005MNRAS.358..101M} or numerical
simulations \citep{2008ApJ...673..657M}.

\subsection{WMAP5 CMB anisotropies}

To calculate CMB temperature and polarisation power- and
cross-spectra we use the publicly available package
CAMB\footnote{http://camb.info} \citep{Lewis:1999bs}.  The likelihood
is evaluated using the public WMAP5
code\footnote{http://lambda.gsfc.nasa.gov}
\citep{WMAP5-Dunkley08}. We include all `standard' components which
are the same for three- and five-year. Those are the low-$\ell$ TT and
TE/EE/BB spectra, the high-$\ell$ TT and TE spectra, and the
point-source TT correction. The low-$\ell$ ($\ell \le 32$ for TT,
$\ell \le 23$ for polarisation) likelihoods
are calculated using Gibbs sampling
\citep{WMAP3-Page07, WMAP5-Dunkley08}. The high-$\ell$ sampling
uses pseudo-$C_\ell$ according to \citet{WMAP3-Hinshaw07}.

In contrast to the published WMAP5 results, we do not include
corrections due to SZ. The SZ amplitude is unconstrained by WMAP5 and
is not degenerate with other parameters \citep{WMAP5-Dunkley08}. A
recent analysis found no biases for WMAP5 when ignoring the
contribution of the thermal SZ effect from clusters \citep{taburet-2008}.

\section{Parameter analysis and cosmological constraints}
\label{sec:constraints}

\subsection{Cosmological model and parameters}

We assume a flat dark-energy cold dark matter cosmology ($w$CDM) with
the parameter vector $(\Omegam, \Omegab, \tau, w, \ns, h,
\sigma_8)$. Clustering of dark energy is not taken into account. For
CMB, instead of the normalisation defined at a scale of 8 Mpc/h,
$\sigma_8$, the parameter which is sampled is $\Delta_{\cal R}^2$, the
curvature perturbations amplitude at the pivot scale $k_0 = 0.002 \,
{\rm Mpc}^{-1}$. As stated in Sect.~\ref{sec:lensing-data} we use the
\citet{2003MNRAS.341.1311S} fitting formula to model the non-linear
power spectrum needed for weak lensing. We do not include a massive
neutrino component to the mass-energy tensor and assume $N_{\rm eff} =
3.04$ as the effective number of massless neutrinos
\citep{2002PhLB..534....8M}, which is the preferred value for WMAP5
\citep{WMAP5-Dunkley08}. \citet{TSUK09} obtained constraints on the
neutrino mass using various probes including CFHTLS-Wide weak lensing,
see also \cite{2008arXiv0810.3572G} and \cite{2009PhRvD..79b3520I}.
Additional parameters are $(a,b,c,c_0)$ from lensing and $(M, \alpha,
\beta, \theta_{10}, \ldots \theta_{17})$ from SN Ia. These parameters
are described in the respective subsections of Sect.~\ref{sec:data};
including systematics we sample a total of 22 parameters.

\subsection{Sampling the parameter space with MCMC}

We use an adaptive Metropolis-Hastings algorithm \citep{Metropolis53,
  Hastings70} to generate Monte Carlo Markov Chains (MCMC) as sample
of the posterior.  The Fisher matrix evaluated at the
maximum-likelihood (ML) parameter serves as initial multi-variate
Gaussian proposal. The ML point is estimated by a conjugate-gradient
search \citep{nr}.  For lensing and SNIa the typical chain length is
100\,000 with an acceptance rate of about 15\%-20\%. Every 1\,000
steps the chain covariance is updated and replaces the previous
proposal. The proposal variance is multiplied by a factor
$2.4/\sqrt{n_{\rm dim}}$ which is optimal for a Gaussian posterior
\citep{Hanson98posteriorsampling, 2005MNRAS.356..925D}.

For WMAP5 on the other hand, we choose a different strategy. With
acceptance rates as stated above we found the posterior not well
sampled and as a consequence the parameter uncertainties largely
underestimated.  We refrain from updating the proposal but kept the
initial guess, which is larger than the Fisher matrix, for the whole
duration of the chain. This choice results in small acceptance rates
of 3\% to 10\% but guarantees an unbiased sampling of the posterior.

The experiments are combined by multiplying the corresponding
likelihoods or, equivalently, by summing up the log-likelihoods. The
confidence intervals are obtained by creating histograms of the
parameter vectors of the chain and estimating regions with 68\%, 95\%
and 99.7\% of the density.

In particular for WMAP5 the calculation of the converged chain is very
time-consuming and takes a few days to a week on a fast multi-core
machine. To overcome this and other problems related to MCMC, such as the
difficulty to assess convergence and the high correlation of a Markov
chain, a new method called Population MonteCarlo
\cite[PMC,][]{CGMR03, CDGMR07} has been developed which is based on adaptive
importance sampling. In a companion paper, we test this method
with simulations and apply it to cosmology posteriors \citep{WK08}.


\subsection{Combined constraints from lensing and CMB}

\begin{figure}[!tb]
  
  \resizebox{0.9\hsize}{!}{
    \includegraphics{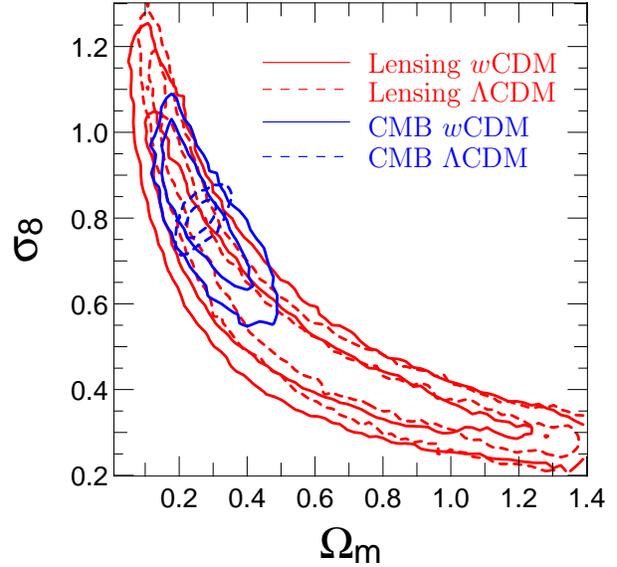}
  }
  
  \caption{68\% and 95\% confidence levels for lensing only, assuming $w$CDM
    (solid red lines), $\Lambda$CDM (dashed red), and CMB only, for $w$CDM
    (solid blue), $\Lambda$CDM (dashed blue curves).}
  \label{fig:w_LCDM}
\end{figure}

The main contribution from cosmic shear (without tomographic
information), in combination with CMB anisotropy measurements, are
constraints on the normalisation $\sigma_8$ \citep{Contaldi03,
  2005AA...429..383T}. For a $\Lambda$CDM model, the degeneracy
direction between $\Omegam$ and $\sigma_8$ is orthogonal between
lensing and CMB. Lensing is sensitive to the amount of structure and
therefore, a high normalisation has to be countered by a smaller
amount of the total matter. On the other hand, an overall increase of
the CMB angular power spectrum by a higher normalisation has to be
compensated by more matter to dampen the peaks.

This picture changes if the dark energy equation-of-state is a free
parameter. Whereas the lensing contours for $\Omegam$ and $\sigma_8$
do not broaden much, the CMB constraints increase dramatically. Most
importantly, they increase along the lensing-direction of
degeneracy, see Fig.~\ref{fig:w_LCDM}. The result is that adding
lensing to CMB data will not improve the constraints in the
$\Omegam$-$\sigma_8$ plane as much as it would do in the framework of
a cosmological constant model.

Lensing without tomography is not very sensitive to $w$, thus leaving
it a free parameter has a minor influence on the other parameters. For
CMB however, strong correlations enter through the angular diameter distance
which CMB measures only at $z=1100$. Those degeneracies (see
Fig.~\ref{fig:all_5LS}) can be broken by adding an additional distance
measurement at low redshift as a lever, e.g.~using BAO or SNIa,
\citep[][]{2003PhRvD..67h3505F, 2003ApJ...594..665B,
  2003ApJ...598..720S}.

\subsection{Combined constraints from lensing, SNIa and CMB}

\begin{table*}
  \caption{CMB, lensing and SNIa in various combinations. The mean and
    68\% marginals are given. For the first four cases systematics are
    ignored, the last column includes all systematics, from both lensing and supernovae (see
    Sect.~\ref{sec:bias_sys}).
  }
  \label{tab:5LS}
  \renewcommand{\arraystretch}{1.5}
  \begin{center}\begin{tabular}{|l|l|l|l|l|l|}\hline
\rule[-3mm]{0em}{8mm}Parameter	 & CMB	 & CMB+Lens	 & CMB+SN	 & CMB+Lens+SN	 & CMB+Lens+SN+sys	\\ \hline\hline
$\Omega_{\textrm{b}}$	 & $0.045^{+0.020}_{-0.016}$	 & $0.041^{+0.016}_{-0.008}$	 & $0.0433^{+0.0028}_{-0.0026}$	 & $0.0432^{+0.0026}_{-0.0023}$	 & $0.0428\pm{0.0029}$	\\ \hline
$\Omega_{\textrm{m}}$	 & $0.262^{+0.099}_{-0.093}$	 & $0.242^{+0.092}_{-0.048}$	 & $0.257^{+0.025}_{-0.023}$	 & $0.253^{+0.018}_{-0.016}$	 & $0.251^{+0.023}_{-0.018}$	\\ \hline
$\tau$	 & $0.087\pm{0.016}$	 & $0.086^{+0.016}_{-0.017}$	 & $0.088^{+0.019}_{-0.016}$	 & $0.088^{+0.019}_{-0.015}$	 & $0.088\pm{0.017}$	\\ \hline
$w$	 & $-1.08^{+0.39}_{-0.53}$	 & $-1.09^{+0.24}_{-0.22}$	 & $-1.025^{+0.071}_{-0.072}$	 & $-1.010^{+0.059}_{-0.060}$	 & $-1.021^{+0.079}_{-0.081}$	\\ \hline
$n_{\textrm{s}}$	 & $0.963^{+0.019}_{-0.014}$	 & $0.961^{+0.014}_{-0.016}$	 & $0.962\pm{0.015}$	 & $0.963^{+0.015}_{-0.014}$	 & $0.963^{+0.014}_{-0.015}$	\\ \hline
$10^9\Delta^2_R$	 & $2.43^{+0.13}_{-0.14}$	 & $2.418^{+0.083}_{-0.110}$	 & $2.43^{+0.12}_{-0.11}$	 & $2.414^{+0.098}_{-0.092}$	 & $2.41\pm{0.11}$	\\ \hline
$h$	 & $0.74^{+0.18}_{-0.12}$	 & $0.754^{+0.096}_{-0.089}$	 & $0.719^{+0.025}_{-0.022}$	 & $0.720^{+0.023}_{-0.021}$	 & $0.723^{+0.027}_{-0.025}$	\\ \hline
$\sigma_8$	 & $0.82^{+0.14}_{-0.15}$	 & $0.819^{+0.061}_{-0.069}$	 & $0.807^{+0.044}_{-0.046}$	 & $0.795^{+0.030}_{-0.027}$	 & $0.798^{+0.037}_{-0.044}$	\\ \hline
\end{tabular}\end{center}

  \renewcommand{\arraystretch}{1}
\end{table*}

\begin{figure}[!tb]

  \resizebox{\hsize}{!}{
    \includegraphics{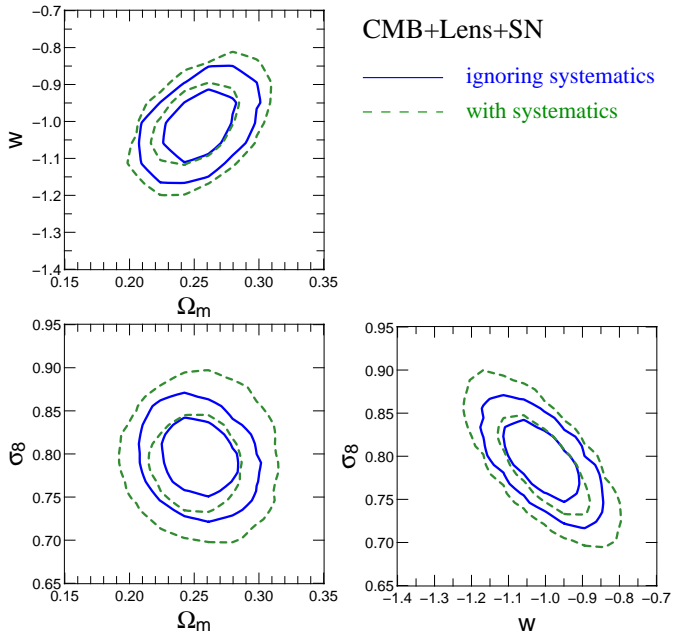}
  }

  \caption{68\% and 95\% confidence levels for the three parameters
    which are affected most by systematics ($\Omegam, w,
    \sigma_8$). Solid (dashed) contours correspond to the case of
    ignoring (including) systematics.}
  \label{fig:nosys_sys}
\end{figure}

\begin{figure*}[!tb]

  \resizebox{\hsize}{!}{
    \includegraphics{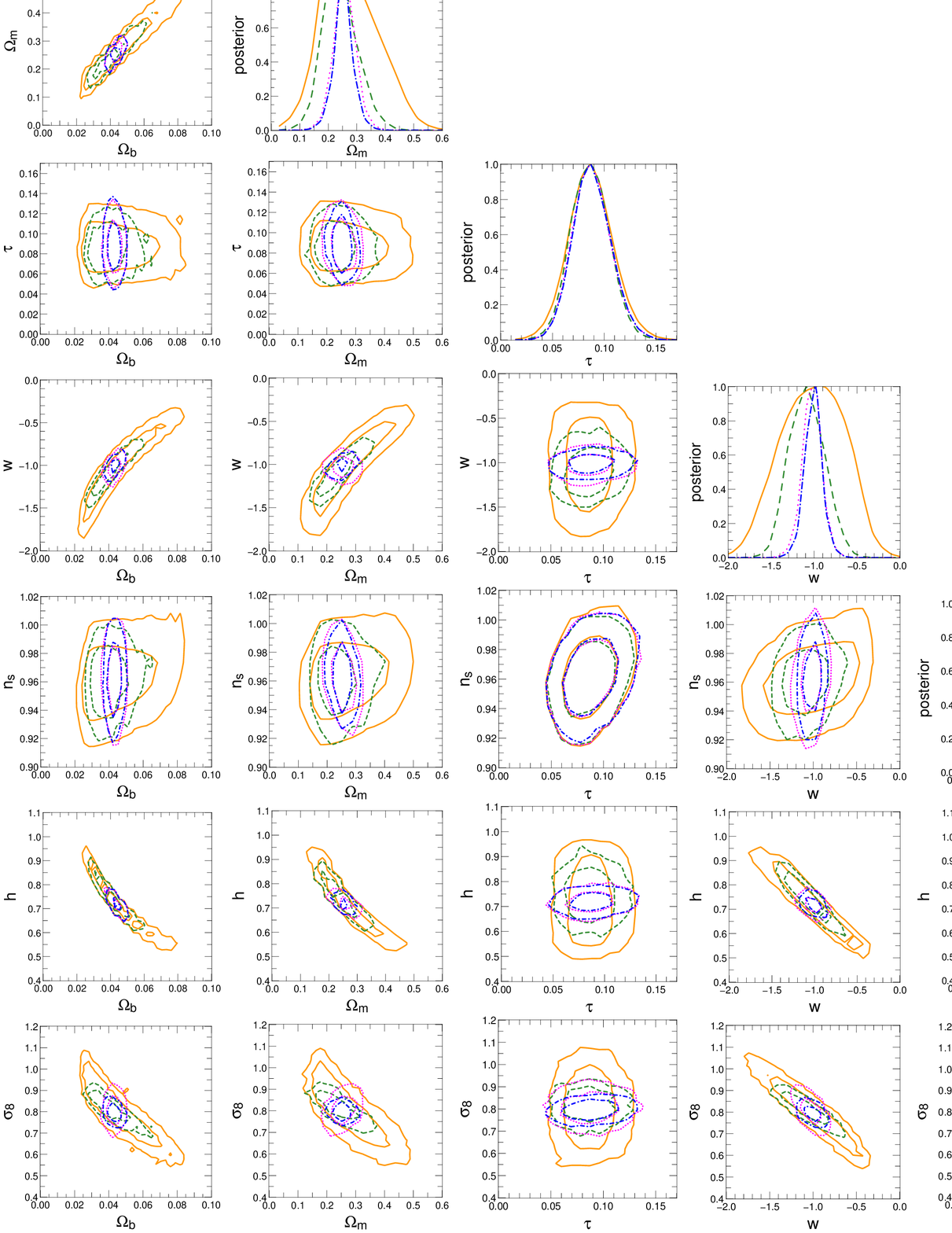}
  }

  \caption{68\% and 95\% confidence levels for CMB (orange, solid
    lines), CMB+lensing (green dashed), CMB+SNIa (magenta dotted) and
    CMB+lensing+SNIa (blue dash-dotted). Systematics are ignored in
    this plot.}
  \label{fig:all_5LS}
\end{figure*}

\begin{figure*}[!tb]
  \resizebox{\hsize}{!}{
    \includegraphics{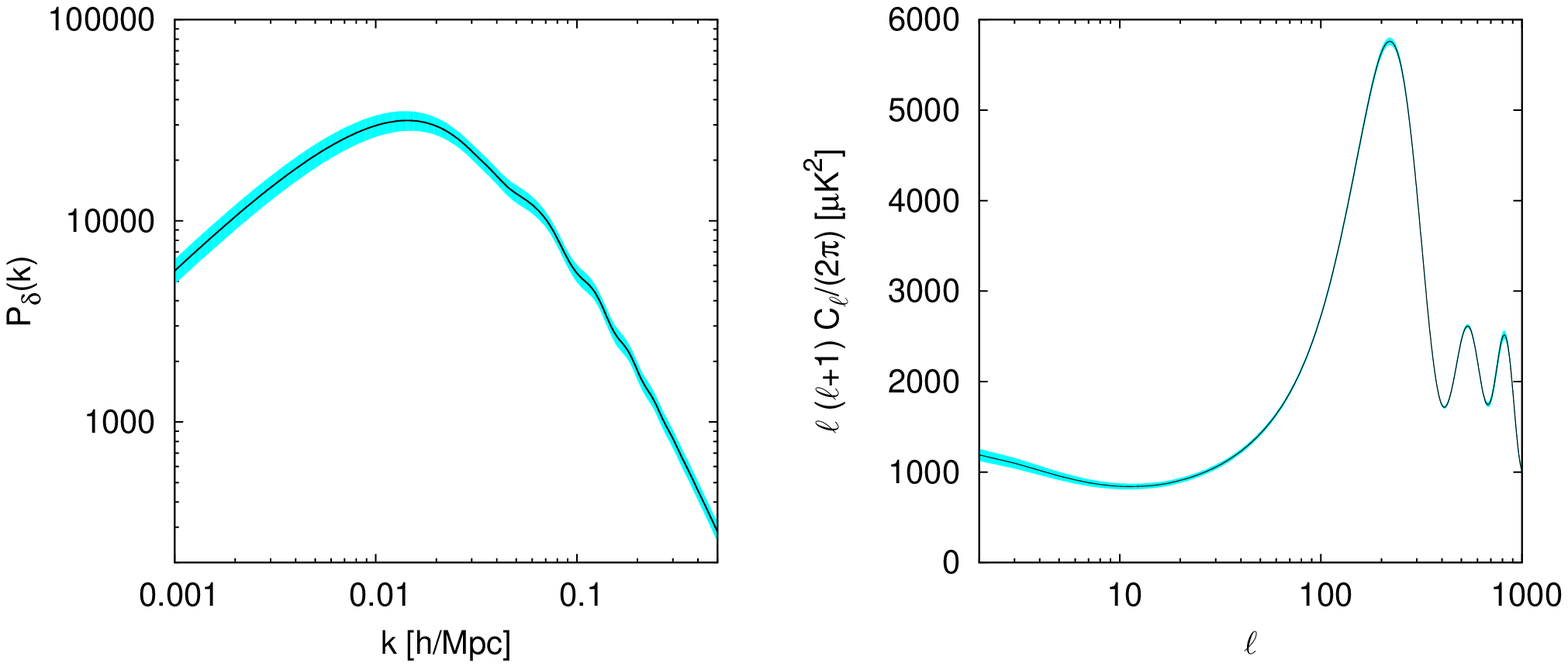}
  }
  \caption{The combined (lensing+SNIa+CMB) best-fit $z=0$ matter power
    spectrum (left panel) and angular power spectrum (right
    panel). The shaded region corresponds to the 68\% uncertainty
    including all systematics. Note that this error does not
    correspond to the uncertainty due to cosmic variance.}
  \label{fig:power_spectra}
\end{figure*}

We will now discuss the main results of this paper. Joint
constraints using lensing, SNIa and CMB are compared for the two cases
with and without taking systematics into account. In the former, both
lensing and supernovae systematics are included. The results are given
in Table \ref{tab:5LS}, Fig.~\ref{fig:nosys_sys} and
  Fig.~\ref{fig:all_5LS}. We show the best-fit angular and 3d power
spectra in Fig.~\ref{fig:power_spectra}. A CAMB parameter file with
our best-fit parameter values is available for
download\footnote{{http://www2.iap.fr/users/kilbinge/params.ini}}.

With the current data, SNIa is more efficient than lensing in helping
decrease uncertainties. CMB+SNIa gives nearly as tight constraints as
CMB+SNIa+lensing for most parameters. The reason is that SNIa data show
different degeneracy directions, in particular for the pair
$\Omegam$-$w$ (see right panel of Fig.~\ref{fig:zeropt2d}). This helps
to pin down $w$ and thus, the main degeneracy for CMB is largely
lifted. As stated in the previous section, this is not the case for
lensing --- lensing without tomography cannot constrain $w$. The
consequence is that even parameters to which SNIa is not sensitive,
e.g.\ $\Omegab$, are very accurately determined for the combination
CMB+SNIa.

Nevertheless, lensing improves constraints from WMAP5 substantially. Some
CMB-related near-degeneracies which arise in the $w$CDM model are
partially lifted in combination with lensing. In particular for
($\Omegam, \sigma_8$), there is a large gain when lensing data is added.

The effect of systematics on the parameters means and errors can be
assessed by comparing the last two columns of Table \ref{tab:5LS}. The
shift of the best-fit values is less than 15\% of the statistical
error in all cases. Including systematics in the analysis increases
the error bars by 10\%-35\%, where $\Omegam$, $w$ and $\sigma_8$ are
affected most.
Varying the lensing redshift for the bias $z_0$ from 1.0 to 0.8
  and 1.2 changes the results by less then one percent.

\section{Discussion}
\label{sec:discussion}

In this paper we combine three different cosmological probes to test a
possible deviation from a cosmological constant: (1) Weak
gravitational lensing as a probe of structure formation and geometry
in the redshift range of about $0.2$ to $0.8$. (2) Supernovae Ia as
standard(isable) candles up to redshift unity. (3) CMB anisotropies supplying
a wealth of information of the recombination era ($z
\approx 1100$) and, to a lesser extend, of the Universe up to the
re-ionisation epoch ($z \approx 6 - 15$). The data sets are (1)
CFHTLS-Wide T0003 (see F08), (2) SNLS first-year (A06) and (3) WMAP
five-year (H09), respectively.

We test models in the context of a flat CDM cosmology with a dark- or
vacuum-energy component with free but constant equation-of-state
parameter $w = p/\rho c^2$. This corresponds to the simplest extension
of the `vanilla' $\Lambda$CDM model which goes beyond a cosmological
constant.

The joint constraints including the modelled systematics in the data
sets are $w = -1.02^{+0.08 +0.14}_{-0.08 -0.16}$ (68\% and 95\%
confidence, respectively). Without taking the systematics into
account, the result is $w = -1.01 \pm 0.06 \pm 0.12$, representing
25\% smaller error bars.  With the current data there is no evidence
for a dynamical dark-energy component not being the cosmological
constant.

Two potential sources of bias in the third-year CFHTLS-Wide lensing
data are scrutinised. One, the measured variations of the shear signal
(aperture-mass dispersion $\langle M_{\rm ap}^2 \rangle$) between
MegaCam pointings are compared to $N$-body simulations. We estimate
the measured fluctuations to be higher than expected by not more than
about 5\% to 15\% on scales below 30 arc minutes. Whereas this might
be a hint of systematics in the data, it is not straightforward to
model and to assess its effect on cosmology. The second issue are
systematics in the shape measurements which seem to lead to an
underestimation of the lensing signal at high redshift. We devised a
very simple model of this potential systematics by multiplying the
lensing efficiency above $z=1$ with a constant $c_0>0$, to mimic the
effect of a decreased measured shear. Marginalising over cosmological
parameters, using weak lensing alone, yields $c_0 = 1.1 \pm 0.6$. All
probes combined do not constrain $c_0$ much better, we find $c_0 =
1.1\pm 0.5$.  Restricting ourselves to $c_0<1$, implying an
underestimation of the lensing signal, increases $\sigma_8$ by about
8\% for a fixed $\Omegam = 0.25$, which is roughly equal to the
statistical error. Therefore, in the framework of this simple model,
the value of $\sigma_8$ might be biased by 8\% if this
effect is ignored (as it has been in F08).

Combining all probes we assess the influence of systematics on the
cosmological results. The contribution of systematics to the total
error budget ranges between 5\% and 40\% (Table \ref{tab:5LS}). The
parameters which are affected most by systematics are $\Omegam,
\sigma_8$ and $w$ for which this contribution is greater than
20\%. Those are the parameters for which weak lensing adds
significantly to the joint constraints.

There are indications of more unaccounted systematics in the lensing
data. Discussions about the origin of those systematics and ways to
remove them are addressed elsewhere (van Waerbeke et al.~in
prep.). Our findings strengthen the confidence in cosmological results
from cosmic shear, presented in this work and earlier, using the same
data, in F08.

Our constraints are slightly tighter than \citet{0407372} who obtained
$w = -0.99^{+0.09 + 0.16}_{-0.09 -0.20}$ using WMAP1, SNIa and, as
probes of structure formation, SDSS galaxy correlations and Ly$\alpha$
forest clustering. Similar constraints were quoted in
\citet{2006ApJ...650....1W}, $w = -1.00^{+0.08 + 0.16}_{-0.08 - 0.17}$
from WMAP3, SNLS and SDSS. Our results are comparable to the ones
stated by \citet{WMAP5-Komatsu08}, $-0.11 < 1+w < 0.14$ (95\%)
stemming from WMAP5, BAO and SNIa. Consistency is also achieved with
the constraints from \citet{2008MNRAS.387.1179M} who combined WMAP5
with the X-ray cluster mass function, cluster baryon fraction and SNIa
to get 
$w = -1.02 \pm 0.06$ (68.3\%).

Earlier results including weak gravitational lensing used less wide
and/or more shallow data. Using the first-year release of the CFHTLS
\cite[T0001,][]{CFHTLSwide}, and with rather tight priors on other
cosmological parameters, an upper bound of $w<-0.8 \, (68\%)$ was
derived. \cite{JBBD06} combined weak lensing from the 75 deg$^2$ CTIO
survey with SNIa and CMB and obtained $w = -0.89^{+0.16}_{-0.21} \,
(95\%)$. The error estimate for the latter includes statistical
uncertainty, PSF systematics and shape measurement calibration biases.

This paper puts particular emphasis on the treatment of systematics in
the data. Known observation-related systematics are parametrised if
possible and included in the analysis together with cosmological
parameters. This allows us to directly quantify the influence of
systematics on cosmology and to find possible correlations. For SNIa,
we calculate the response of the distance modulus to fluctuations of
the photometry zero-points. Ignoring these error sources leads to an
underestimation of the parameter errors by 10\%. The parameter values
itself are biased by a fraction of 10\% to 20\% of the statistical
uncertainty.

\section{Outlook}
\label{sec:outlook}

For future, high-precision experiments it will be of great importance
to understand and control systematic effects.
In particular for exotic models, such as time-varying
dark-energy or modifications of GR, more and more subtle influences of
systematics mimicking a signal have to be excluded.
This calls for
combined analyses of cosmological and systematics parameters.

Weak lensing serves as an important and independent probe of $\sigma_8$.
The measurement of the lensing skewness will allow us
to lift degeneracies between parameters, in particular between
$\Omegam$ and $\sigma_8$ \citep{KS05}. Power and bi-spectrum
tomography can be used to constrain the time-evolution of the
dark-energy eos parameter \citep{2004MNRAS.348..897T}. In addition, as
a complementary and nearly independent probe of the non-Gaussian,
high-density regime of the large-scale structure, weak lensing cluster
counts can improve dark-energy parameter constraints
\citep{2007NJPh....9..446T}.

For lensing, systematic effects which have to be taken into account
for future analyses include measurement errors (galaxy shapes, PSF
correction, photo-$z$'s), astrophysical sources (intrinsic alignments,
source clustering) and theoretical uncertainties (non-linear and
baryonic physics). The modelling of all those effects, as suggested
recently by \citet{2008arXiv0808.3400B}, leads to a huge number of
parameters, on the order of several dozens to hundreds, depending on
the number of redshift bins. In case of simultaneous
determination of cosmic shear and intrinsic alignments a high number of
redshift bins is required \citep{2007NJPh....9..444B,2008AA...488..829J}.

Already with the present data, more general models of dark energy can
be constrained. A wide variety of such models are tested with recent
SNIa data \citep{2008arXiv0807.1108R}. Because dark energy and
modified gravity can influence distances in a non-distinguishable way,
it is important to include measures of the growth of structure.
A survey like the CFHTLS-Wide reaches out into the linear
regime and is less prone to small-scale uncertainties due to baryonic and
non-linear physics, which makes such a survey an excellent probe of the
present acceleration of the Universe.

\begin{acknowledgements}

We acknowledge the use of the Legacy Archive for Microwave Background
Data Analysis (LAMBDA). Support for LAMBDA is provided by the NASA
Office of Space Science. We thank the \textsc{Terapix} group for support and
computational facilities. We acknowledge the CFHTLS lensing
systematics collaboration for uncovering the weak-lensing systematics
that are parametrised and marginalised over in this analysis. We thank
O.~Capp\'e, J.-F.~Cardoso and O.~Dor\'e for helpful discussions, and
E.~Bertin, S.~Prunet and B.~Rowe for valuable comments on the
manuscript. We also would like to thank the anonymous referee for helpful
suggestions which improved the paper.
MK is supported by the CNRS ANR ``ECOSSTAT'', contract
number ANR-05-BLAN-0283-04. IT and LF acknowledge the support of the
European Commission Programme 6-th framework, Marie Curie Training and
Research Network ``DUEL'', contract number MRTN-CT-2006-036133.

\end{acknowledgements}

\bibliographystyle{aa} \bibliography{astro}

\end{document}